\documentclass[australian,showpacs,showkeys,superscriptaddress]{revtex4-1}
\usepackage[T1]{fontenc}
\usepackage[latin9]{inputenc}
\usepackage{textcomp}
\usepackage{amsmath}
\usepackage{graphicx}
\usepackage{babel}
\begin{document}

\title{Generalised balance equations for charged particle transport via
localised and delocalised states: Mobility, generalised Einstein relations
and fractional transport}

\author{Peter W. \surname{Stokes}}
\email[Electronic address: ]{peter.stokes@my.jcu.edu.au}

\affiliation{College of Science and Engineering, James Cook University, Townsville,
QLD 4811, Australia}

\author{Bronson \surname{Philippa}}

\affiliation{College of Science and Engineering, James Cook University, Cairns,
QLD 4870, Australia}

\author{Daniel Cocks}

\affiliation{College of Science and Engineering, James Cook University, Townsville,
QLD 4811, Australia}

\author{Ronald D. \surname{White}}

\affiliation{College of Science and Engineering, James Cook University, Townsville,
QLD 4811, Australia}
\email[Electronic address: ]{ronald.white@jcu.edu.au}

\begin{abstract}
A generalised phase-space kinetic Boltzmann equation for highly non-equilibrium
charged particle transport via localised and delocalised states is
used to develop continuity, momentum and energy balance equations,
accounting explicitly for scattering, trapping/detrapping and recombination
loss processes. Analytic expressions detail the effect of these microscopic
processes on the mobility and diffusivity. Generalised Einstein relations
(GER) are developed that enable the anisotropic nature of diffusion
to be determined in terms of the measured field-dependence of the
mobility. Interesting phenomena such as negative differential conductivity
and recombination heating/cooling are shown to arise from recombination
loss processes and the localised and delocalised nature of transport.
Fractional transport emerges naturally within this framework through
the appropriate choice of divergent mean waiting time distributions
for localised states, and fractional generalisations of the GER and
mobility are presented. Signature impacts on time-of-flight current
transients of recombination loss processes via both localised and
delocalised states are presented.
\end{abstract}


\maketitle

\section{\label{sec:Introduction}Introduction}

Dispersive transport is characterised by a mean squared displacement
that scales sublinearly with time \citep{Metzler1999}. Physically,
this fundamentally slower transport can arise due to the presence
of trapped (localised) states, causing the temporary immobilisation
of particles \citep{Scher1975}. Some examples include charge carrier
trapping in local imperfections of organic semiconductors \citep{Scher1975,Sibatov2007},
electron trapping in bubble states within liquid neon and liquid helium
\citep{Mauracher2014,Borghesani2002,Sakai1992}, ion trapping in liquid
xenon \citep{Hilt1994,Hilt1994a,Hilt1994b,Schmidt2005}, positronium
trapping in bubbles \citep{Stepanov2012,Stepanov2002,Charlton2001}
and positron annihilation on induced clusters \citep{Colucci2011}.
Trapped states also exist in organic-inorganic metal-halide perovskites
and influence the delocalised nature of transport in these materials
\citep{Wetzelaer2015}. The combined localised/delocalised nature
of charged transport occurring in many materials warrants the development
of a new transport theory to treat and explore the problem, and this
represents the theme of our program.

In our previous work \citep{Stokes2016} we explored a generalised
phase-space kinetic model for charged particle transport that considered
separate collisional, trapping/detrapping and recombination loss processes.
This model takes the form of a generalised Boltzmann equation with
operators that describe each process. Rather than performing a direct
solution of Boltzmann's equation, as considered in \citep{Stokes2016},
in this study we embrace a more physical insight and explore the relationships
between the measured macroscopic transport properties and the underlying
microscopic processes (as determined by the appropriate collision
frequencies). This is a philosophy that has been adopted in swarm
physics, and now is routinely applied in a variety of fields including
low-temperature plasma physics \citep{Mason1988,Robson1986,Robson2005,Dujko2013,Dujko2015},
positron physics \citep{White2011,Boyle2012,Robson2015}, liquid particle
detectors \citep{Boyle2015,Boyle2016} and radiation damage \citep{Ness2012,White2014,DeUrquijo2014}.

For gaseous systems, or those where transport occurs through delocalised
states, there exists a wealth of literature that explores relationships
between experimentally measurable transport properties, and links
the underlying microscopic physics to the macroscopic through simple
analytic expressions. In fact, transport properties were initially
used as the means to indirectly measure scattering cross-sections
and their energy dependence. In this study, we aim to generalise many
existing results for such systems and explore the impact of localised
(trapped) states and loss/recombinations on (i) the mobility, (ii)
the Wannier energy relation \citep{Wannier1953}, which relates the
mean energy of the charged particles to the mobility, and (iii) the
Einstein relations \citep{Robson1976,Robson1984} which relate the
mobility to the diffusivity and enable the quantification of the anisotropic
nature of diffusion. Using these we postulate the existence of a number
of new phenomena, including trap-induced particle heating/cooling
and trap-induced negative differential conductivity (NDC), the origin
of which differs significantly from that in which transport is delocalised.
Criteria on the various collision, trapping and loss frequencies are
presented for the occurrence of such phenomena. 

In Sec. \ref{sec:Model} of this paper we present a generalised Boltzmann
equation with energy-dependent process rates for collisions, trapping
and recombination. We explore the signature impact of recombination
loss processes in both the delocalised and localised states on the
time-of-flight current transients in Sec. \ref{sec:Charge}. In Sec.
\ref{sec:Balance}, balance equations are formed for particle continuity,
momentum and energy, via the appropriate moments of the generalised
Boltzmann equation, which are also used to develop expressions for
mobility, mean energy and diffusivity. Phenomena such as heating/cooling,
NDC, and generalised Einstein relations (GER) are explored in Secs.
\ref{sec:Phenomena}\textendash \ref{sec:GER}. In Sec. \ref{sec:Fractional},
the fractional transport equivalents of the above are considered including
fractional GER, while in Sec. \ref{sec:Conclusion}, we present conclusions
and outline some possible avenues for future work.

\section{\label{sec:Model}Extended phase-space model}

\begin{figure}
\includegraphics{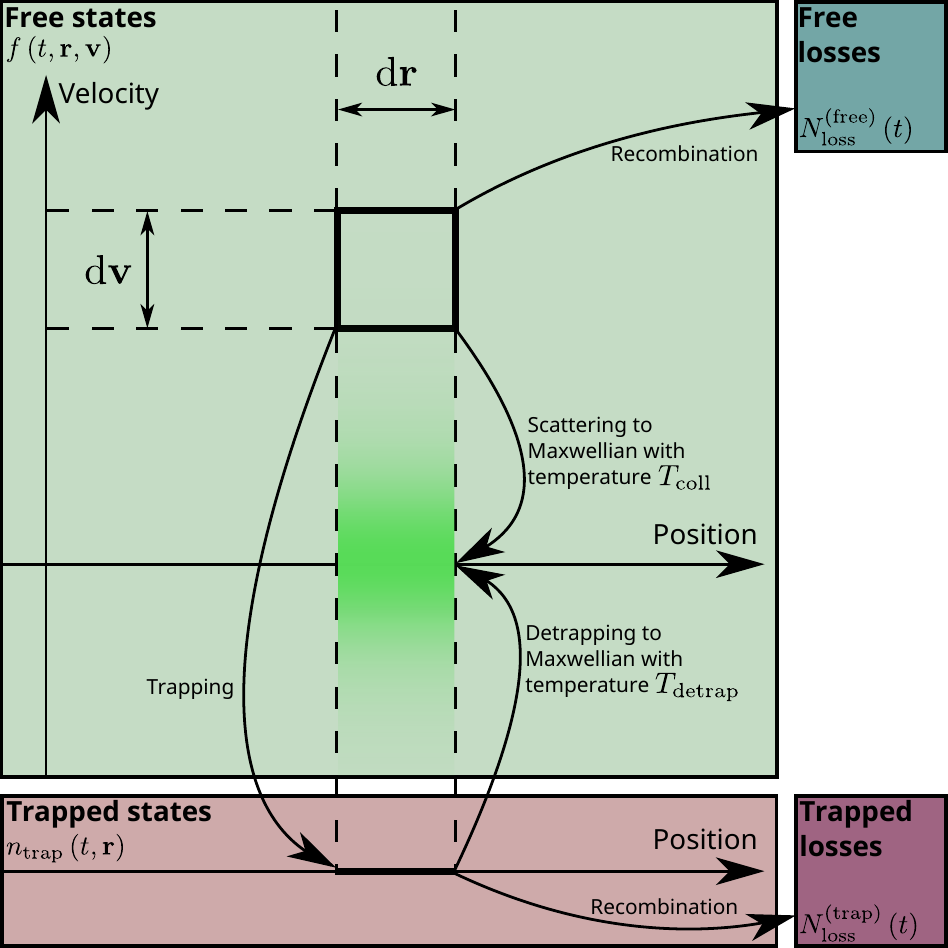}
\centering{}\caption{\label{fig:phaseSpace}Phase-space diagram illustrating the collision,
trapping, detrapping and recombination processes. (Source: \citep{Stokes2016})}
\end{figure}

In this section, we consider a generalisation of the kinetic model
presented in Eq. (1) from reference \citep{Stokes2016} that describes
the processes of collisions, trapping and recombination, as depicted
in Fig. \ref{fig:phaseSpace}. Specifically, we make processes selective
of particle energy $\epsilon\equiv\frac{1}{2}mv^{2}$. This results
in a free particle phase-space distribution function $f\left(t,\mathbf{r},\mathbf{v}\right)$,
defined by the generalised Boltzmann equation 
\begin{eqnarray}
\left(\frac{\partial}{\partial t}+\mathbf{v}\cdot\frac{\partial}{\partial\mathbf{r}}+\frac{e\mathbf{E}}{m}\cdot\frac{\partial}{\partial\mathbf{v}}\right)f\left(t,\mathbf{r},\mathbf{v}\right) & = & -\nu_{\mathrm{coll}}\left(\epsilon\right)f\left(t,\mathbf{r},\mathbf{v}\right)+n\left(t,\mathbf{r}\right)\left\langle \nu_{\mathrm{coll}}\left(\epsilon\right)\right\rangle \tilde{w}_{\mathrm{coll}}\left(v\right)\nonumber \\
 &  & -\nu_{\mathrm{trap}}\left(\epsilon\right)f\left(t,\mathbf{r},\mathbf{v}\right)+\Phi\left(t\right)\ast\left[n\left(t,\mathbf{r}\right)\left\langle \nu_{\mathrm{trap}}\left(\epsilon\right)\right\rangle \right]\tilde{w}_{\mathrm{detrap}}\left(v\right)\nonumber \\
 &  & -\nu_{\mathrm{loss}}^{\left(\mathrm{free}\right)}\left(\epsilon\right)f\left(t,\mathbf{r},\mathbf{v}\right),\label{eq:boltzmannEquation}
\end{eqnarray}
which describes particles of charge $e$ and mass $m$ in the presence
of an applied electric field $\mathbf{E}$. Here, the energy-dependent
process rates for collisions, trapping and recombination losses are
respectively denoted $\nu_{\mathrm{coll}}\left(\epsilon\right)$,
$\nu_{\mathrm{trap}}\left(\epsilon\right)$, $\nu_{\mathrm{loss}}^{\left(\mathrm{free}\right)}\left(\epsilon\right)$,
$\ast$ denotes a time convolution $\left\langle \cdot\right\rangle $
denotes an average over velocity space:
\begin{align}
\left\langle \psi\left(\mathbf{v}\right)\right\rangle  & \equiv\frac{1}{n\left(t,\mathbf{r}\right)}\int\mathrm{d}\mathbf{v}\,f\left(t,\mathbf{r},\mathbf{v}\right)\psi\left(\mathbf{v}\right),\label{eq:velocityAverage}
\end{align}
where the free particle number density is defined $n\left(t,\mathbf{r}\right)\equiv\int\mathrm{d}\mathbf{v}f\left(t,\mathbf{r},\mathbf{v}\right)$.
Collisions are described above by the Bhatnagar-Gross-Krook (BGK)
collision operator \citep{Bhatnagar1954}, while trapping and detrapping
is described by a BGK-type model with a delay for the duration of
each localised state \citep{Philippa2014}. This delay is sampled
from the effective waiting time distribution \citep{Stokes2016} 
\begin{equation}
\Phi\left(t\right)\equiv\mathrm{e}^{-\nu_{\mathrm{loss}}^{\left(\mathrm{trap}\right)}t}\phi\left(t\right),
\end{equation}
defined in terms of a distribution of trapping times $\phi\left(t\right)$
and weighted by an exponential decay term that describes the recombination
of trapped particles at the rate $\nu_{\mathrm{loss}}^{\left(\mathrm{trap}\right)}$
\citep{Stokes2016}. Note that, unlike the free particle process rates,
this recombination rate is \textit{not} a function of energy as trapped
particles are localised in space.

The processes of scattering and detrapping are taken to be isotropic
and to occur according to Maxwellian velocity distributions. Specifically,
we introduce
\begin{align}
\tilde{w}_{\mathrm{coll}}\left(v\right) & \equiv\frac{\nu_{\mathrm{coll}}\left(\epsilon\right)w\left(\alpha_{\mathrm{coll}},v\right)}{\int\mathrm{d}\mathbf{v}\nu_{\mathrm{coll}}\left(\epsilon\right)w\left(\alpha_{\mathrm{coll}},v\right)},\\
\tilde{w}_{\mathrm{detrap}}\left(v\right) & \equiv\frac{\nu_{\mathrm{trap}}\left(\epsilon\right)w\left(\alpha_{\mathrm{detrap}},v\right)}{\int\mathrm{d}\mathbf{v}\nu_{\mathrm{trap}}\left(\epsilon\right)w\left(\alpha_{\mathrm{detrap}},v\right)},
\end{align}
where the Maxwellian velocity distribution of temperature $T$ is
defined
\begin{eqnarray}
w\left(\alpha,v\right) & \equiv & \left(\frac{\alpha^{2}}{2\pi}\right)^{\frac{3}{2}}\exp\left(-\frac{\alpha^{2}v^{2}}{2}\right),\label{eq:maxwellianDefinition}\\
\alpha^{2} & \equiv & \frac{m}{k_{\mathrm{B}}T},\label{eq:maxwellianTemperature}
\end{eqnarray}
where $k_{\mathrm{B}}$ is the Boltzmann constant.

As stated, this model is very general and requires the precise specification
of atomic and molecular details to properly define the process frequencies.
In practice, this is usually achieved by using cross-section data
in the relationship $\nu\left(\epsilon\right)\equiv n_{0}v\sigma\left(\epsilon\right),$
where $n_{0}$ is the number density of the background medium and
$\sigma\left(\epsilon\right)$ is the cross-section corresponding
to the process of frequency $\nu\left(\epsilon\right)$.

Similar to the description of free particles by Eq. \eqref{eq:boltzmannEquation},
trapped particles can be described by a distribution function in configuration
space $n_{\mathrm{trap}}\left(t,\mathbf{r}\right)$, defined by the
continuity equation
\begin{eqnarray}
\frac{\partial}{\partial t}n_{\mathrm{trap}}\left(t,\mathbf{r}\right) & = & \left(1-\Phi\left(t\right)\ast\right)\left[n\left(t,\mathbf{r}\right)\left\langle \nu_{\mathrm{trap}}\left(\epsilon\right)\right\rangle \right]\nonumber \\
 &  & -\nu_{\mathrm{loss}}^{\left(\mathrm{trap}\right)}n_{\mathrm{trap}}\left(t,\mathbf{r}\right).
\end{eqnarray}
Lastly, the number of particles lost to recombination can also be
counted
\begin{eqnarray}
\frac{\mathrm{d}}{\mathrm{d}t}N_{\mathrm{\mathrm{loss}}}^{\left(\mathrm{free}\right)}\left(t\right) & = & \left\langle \left\langle \nu_{\mathrm{loss}}^{\left(\mathrm{free}\right)}\left(\epsilon\right)\right\rangle \right\rangle N\left(t\right),\\
\frac{\mathrm{d}}{\mathrm{d}t}N_{\mathrm{\mathrm{loss}}}^{\left(\mathrm{trap}\right)}\left(t\right) & = & \nu_{\mathrm{loss}}^{\left(\mathrm{trap}\right)}N_{\mathrm{trap}}\left(t\right),
\end{eqnarray}
where $\left\langle \left\langle \cdot\right\rangle \right\rangle $
denotes an average over phase-space
\begin{equation}
\left\langle \left\langle \psi\right\rangle \right\rangle \equiv\frac{1}{N\left(t\right)}\int\mathrm{d}\mathbf{r}\int\mathrm{d}\mathbf{v}\,f\left(t,\mathbf{r},\mathbf{v}\right)\psi,
\end{equation}
and free and trapped particle numbers are respectively defined
\begin{eqnarray}
N\left(t\right) & \equiv & \int\mathrm{d}\mathbf{r}\,n\left(t,\mathbf{r}\right),\\
N_{\mathrm{trap}}\left(t\right) & \equiv & \int\mathrm{d}\mathbf{r\,}n_{\mathrm{trap}}\left(t,\mathbf{r}\right).
\end{eqnarray}

\section{\label{sec:Charge}Time-of-flight current transients}

In practice, charged particle transport properties can be quantified
using a time-of-flight experiment, where the transit time through
a material for a pulse of charge carriers is found by measuring the
corresponding current. In this section, we explore the impact that
recombination losses of both delocalised and localised particles has
on time-of-flight current transients. We consider the analytical current
in a time-of-flight experiment for a material of thickness $L$ situated
between two plane-parallel electrodes. As this geometry is one-dimensional,
the charge carrier number density $n\left(t,x\right)$ is defined
by the generalised diffusion equation derived in \citep{Stokes2016},
which is rewritten here:
\begin{equation}
\left\{ \frac{\partial}{\partial t}+\nu_{\mathrm{trap}}\left[1-\Phi\left(t\right)\ast\right]+\nu_{\mathrm{loss}}^{\left(\mathrm{free}\right)}\right\} n+W\frac{\partial n}{\partial x}-D\frac{\partial^{2}n}{\partial x^{2}}=0,\label{eq:Generaliseddiffusion}
\end{equation}
where $W$ is the drift velocity and $D$ is the diffusion coefficient.
This diffusion equation can be derived directly from the generalised
Boltzmann equation \eqref{eq:boltzmannEquation}, where the constant
process frequencies can be interpreted as velocity averages of the
energy-dependent frequencies introduced in the previous section, $\nu\equiv\left\langle \nu\left(\epsilon\right)\right\rangle $
. From the number density, the current in a time-of-flight experiment
can be found as the spatially averaged flux \citep{Philippa2011}:
\begin{equation}
j\left(t\right)=e\frac{\partial}{\partial t}\int_{0}^{L}\left(\frac{x}{L}-1\right)n\left(t,x\right)\mathrm{d}x.
\end{equation}
For an impulse initial condition, $n\left(0,x\right)=N\left(0\right)\delta\left(x-x_{0}\right)$,
and perfectly absorbing boundaries, $n\left(t,0\right)=n\left(t,L\right)=0$,
we can proceed as in \citep{Philippa2014} to write this current in
Laplace space:
\begin{equation}
j\left(p\right)=eN\left(0\right)\frac{W}{L\tilde{p}}\left\{ 1-\mathrm{e}^{-\lambda x_{0}}\left[\mathrm{e}^{-\beta x_{0}}+\frac{\sinh\left(\beta x_{0}\right)}{\sinh\left(\beta L\right)}\left(\mathrm{e}^{\lambda L}-\mathrm{e}^{-\beta L}\right)\right]\right\} ,\label{eq:tofCurrent}
\end{equation}
where
\begin{eqnarray}
\tilde{p} & \equiv & p+\nu_{\mathrm{trap}}\left[1-\Phi\left(p\right)\right]+\nu_{\mathrm{loss}}^{\left(\mathrm{free}\right)},\\
\lambda & \equiv & \frac{W}{2D},\\
\beta & \equiv & \sqrt{\frac{\tilde{p}}{D}+\lambda^{2}},
\end{eqnarray}
and the Laplace transform of time, $t\rightarrow p$, is denoted $f\left(p\right)\equiv\mathcal{L}f\left(t\right)\equiv\int_{0}^{\infty}\mathrm{d}t\,\mathrm{e}^{-pt}f\left(t\right)$.
Note that the trapped carrier recombination rate arises here through
the term $\Phi\left(p\right)\equiv\phi\left(p+\nu_{\mathrm{loss}}^{\left(\mathrm{trap}\right)}\right).$

\begin{figure}
\includegraphics{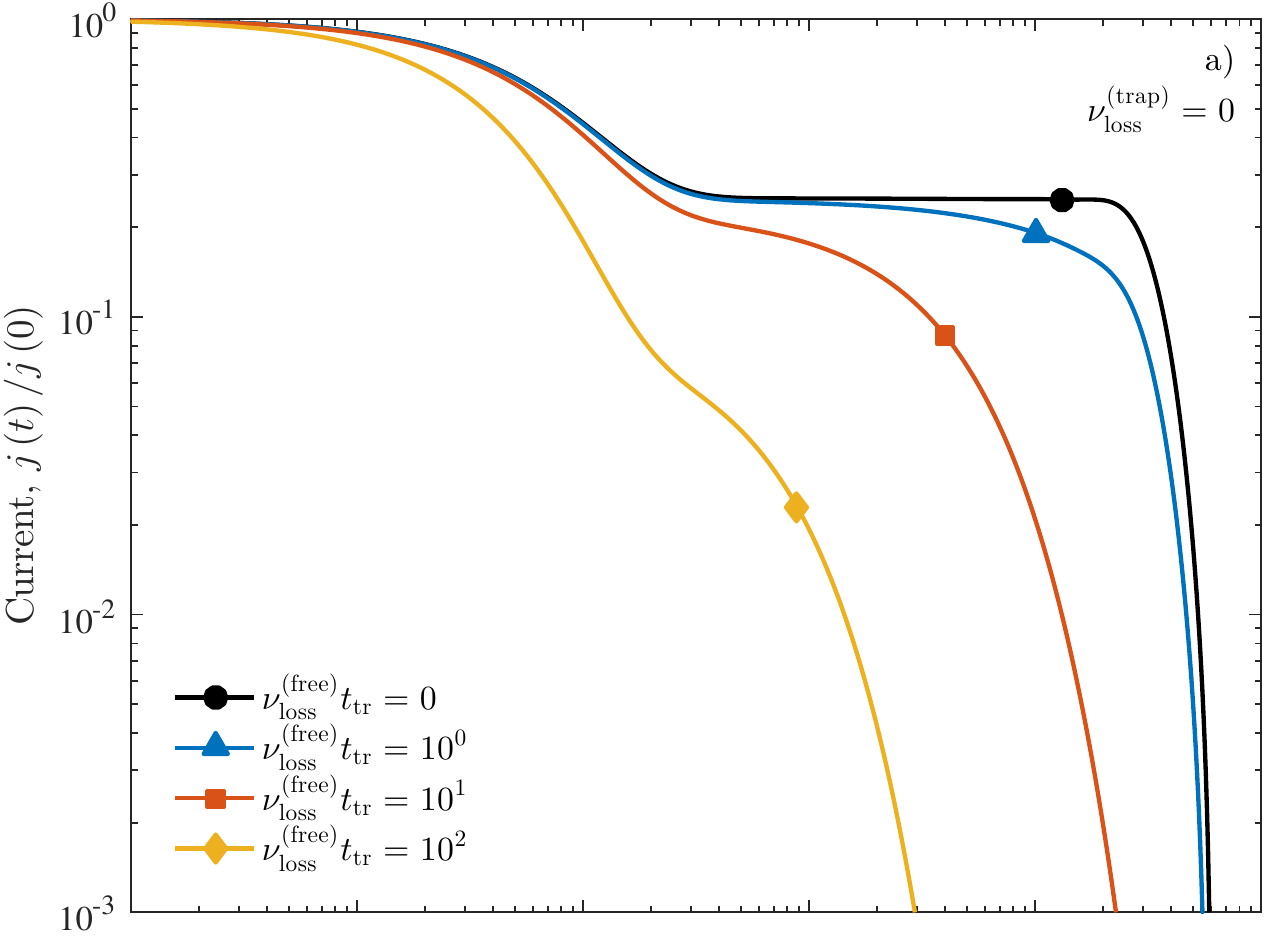}

\includegraphics{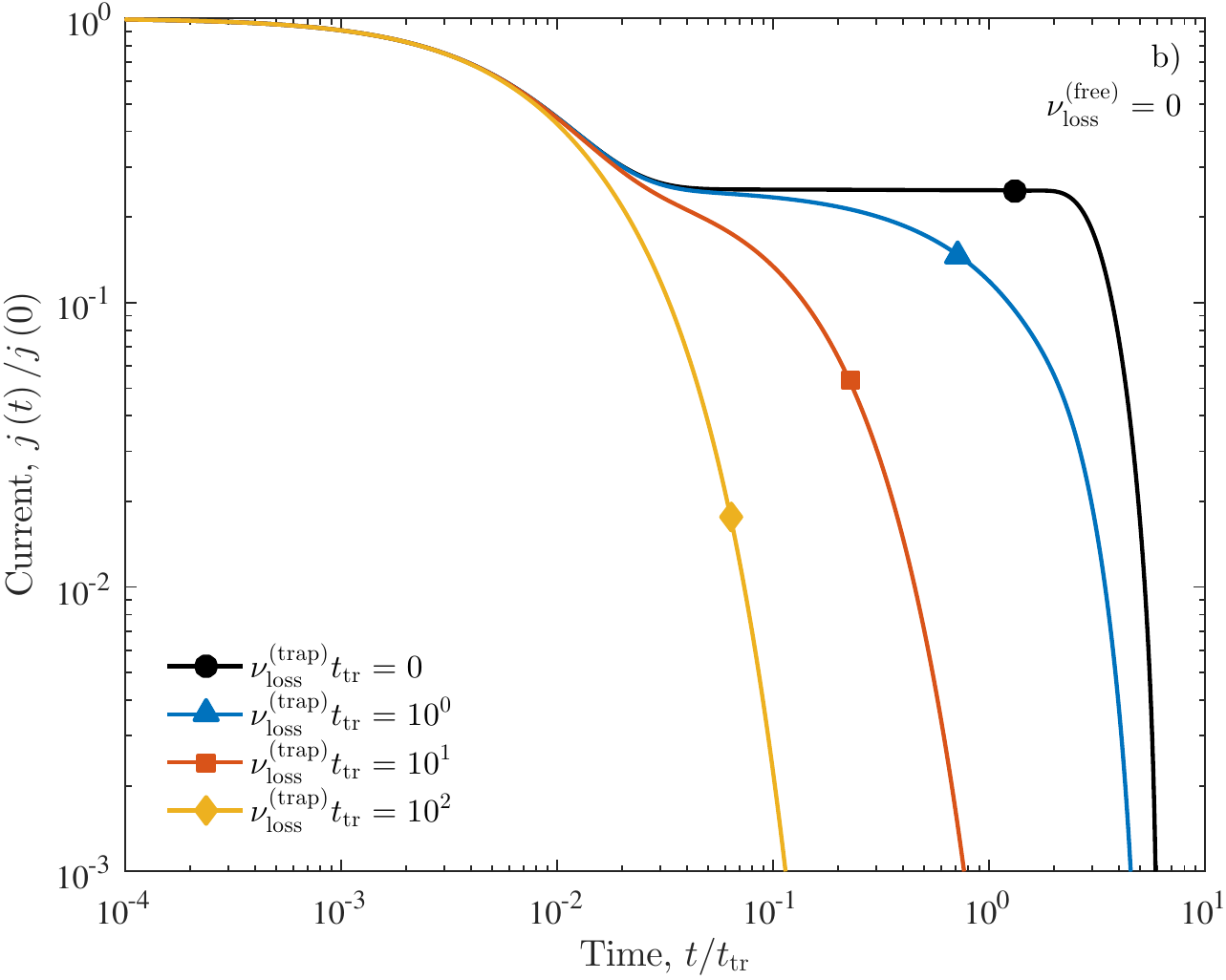}

\caption{\label{fig:current}The impact of free and trapped particle recombination
on current transients for an ideal time-of-flight experiment as modelled
by Eq. \eqref{eq:tofCurrent}. Nondimensionalisation has been performed
using the material thickness $L$, trap-free transit time, $t_{\mathrm{tr}}\equiv L/W$,
and the initial current $j\left(0\right)=eN\left(0\right)/t_{\mathrm{tr}}$.
For these plots we define the diffusion coefficient, $Dt_{\mathrm{tr}}/L^{2}=0.02$,
the initial impulse location, $x_{0}/L=1/3$, and the trapping rate,
$\nu_{\mathrm{trap}}t_{\mathrm{tr}}=10^{2}$. We choose an exponential
distribution of trapping times, $\phi\left(t\right)=\nu_{\mathrm{detrap}}\mathrm{e^{-\nu_{\mathrm{detrap}}t}}$,
with the mean trapping time chosen as $\left(\nu_{\mathrm{detrap}}t_{\mathrm{tr}}\right)^{-1}=0.03$.}
\end{figure}

We consider the explicit effect that free and trapped particle recombination
rates have on the current transient in a time-of-flight experiment
in Fig. \ref{fig:current} by plotting Eq. \eqref{eq:tofCurrent}
for the current, keeping the effects of mobility (drift velocity)
and diffusion constant. A system of units is chosen that uses the
material thickness $L$ and the trap-free transit time, defined as
$t_{\mathrm{tr}}\equiv L/W$. In this system of units, the drift velocity
is equal to unity. We specify the diffusion coefficient to be $Dt_{\mathrm{tr}}/L^{2}=0.02$,
the initial impulse is set to occur at $x_{0}/L=1/3$ and the trapping
rate is made large so as trap-based effects can occur within the transit
time, $\nu_{\mathrm{trap}}t_{\mathrm{tr}}=10^{2}$. For trapping times,
an exponential distribution is considered, $\phi\left(t\right)=\nu_{\mathrm{detrap}}\mathrm{e}^{-\nu_{\mathrm{detrap}}t}$,
with a mean trapping time of $\left(\nu_{\mathrm{detrap}}t_{\mathrm{tr}}\right)^{-1}=0.03$.

In Fig. \ref{fig:current}, the recombination-free current transient
is included in black as a reference. This transient has a number of
notable regimes. At early times, the current is still close to unity
as no processes have had a chance to affect it greatly. What then
follows is a decrease in current as free charge carriers enter traps.
This decrease is temporary, however, and eventually the current plateaus
as a transient equilibrium arises between free and trapped particles.
The value of the current at this plateau is numerically equal to the
proportion of free particles at the equilibrium, $\nu_{\mathrm{detrap}}/\left(\nu_{\mathrm{detrap}}+\nu_{\mathrm{trap}}\right)=0.25\approx10^{-0.6}$.
Finally, the last of the free particles extract causing the remaining
filled traps to gradually exhaust and the system to leave equilibrium.

Fig. \ref{fig:current}a) considers an increasing free particle recombination
rate, $\nu_{\mathrm{loss}}^{\left(\mathrm{free}\right)}$, without
any trapped particle recombination, $\nu_{\mathrm{loss}}^{\left(\mathrm{trap}\right)}=0$.
It can be seen that the free particle losses start decreasing the
current at roughly the characteristic time for free particle recombination,
$\left(\nu_{\mathrm{loss}}^{\left(\mathrm{free}\right)}t_{\mathrm{tr}}\right)^{-1}$.
Because free particles are being lost, an equilibrium is not established
as in the recombination-free case. However, detrapping events do still
cause a slowing in the descent of the current.

Fig. \ref{fig:current}b) considers an increasing trapped particle
recombination rate, $\nu_{\mathrm{loss}}^{\left(\mathrm{trap}\right)}$,
without any free particle recombination, $\nu_{\mathrm{loss}}^{\left(\mathrm{free}\right)}=0$.
Trap-based recombination can only affect the current via detrapping
events and so we do not see a decrease in the current until at least
the characteristic time for trapping, $\left(\nu_{\mathrm{trap}}t_{\mathrm{tr}}\right)^{-1}=10^{-2}$.
Similar to Fig. \ref{fig:current}a), an equilibrium cannot be established
here due to the constant loss of trapped particles. Unlike Fig. \ref{fig:current}a),
however, detrapping events have a diminishing contribution to the
current as increasing trap-based recombination also increases the
probability that trapped particles recombine instead of detrapping.

In practice, time-of-flight current transients will be measured in
experiments. These current traces will be fitted to solutions of the
generalised diffusion equation \eqref{eq:Generaliseddiffusion}, which
enable the transport coefficients (drift velocity $W,$ diffusion
coefficient $D$), various rates $\nu$ and the waiting time distribution
$\phi$ to be determined empirically. In the remainder of this study,
we are focussed on understanding the relationship between the various
microscopic scattering and trapping processes (as determined by the
relevant scattering, trapping and loss collision frequencies and their
dependence on energy, and waiting time distributions) and the transport
coefficients and properties. Furthermore, we will explore relationships
between the transport coefficients/properties e.g. Wannier energy
relation which links the mean energy and the mobility, and the generalised
Einstein relations which link mobility and diffusivity.

\section{\label{sec:Balance}Balance equations}

A knowledge of the full free particle phase-space distribution, $f(t,\mathbf{r},\mathbf{v})$,
defined by the generalised Boltzmann equation \eqref{eq:boltzmannEquation},
is often not required to analyse and interpret experiment. A computationally
economical and more physically appealing alternative is to solve for
average quantities directly, through solution of the appropriate fluid
or velocity moment equations. In what follows, we form these moment
equations by evaluating velocity averages of the phase-space distribution
function, thus grounding them physically through the generalised Boltzmann
equation.

From the Boltzmann equation \eqref{eq:boltzmannEquation}, we show
most generally that the average of a velocity functional $\psi\left(\mathbf{v}\right)$
satisfies the differential equation
\begin{eqnarray}
\frac{\partial}{\partial t}n\left\langle \psi\right\rangle +\frac{\partial}{\partial\mathbf{r}}\cdot n\left\langle \mathbf{v}\psi\right\rangle -\frac{e\mathbf{E}}{m}\cdot n\left\langle \frac{\partial\psi}{\partial\mathbf{v}}\right\rangle  & = & -n\left\langle \psi\nu_{\mathrm{coll}}\left(\epsilon\right)\right\rangle +n\left\langle \nu_{\mathrm{coll}}\left(\epsilon\right)\right\rangle \left\langle \psi\right\rangle _{\mathrm{coll}}\nonumber \\
 &  & -n\left\langle \psi\nu_{\mathrm{trap}}\left(\epsilon\right)\right\rangle +\Phi\left(t\right)\ast\left(n\left\langle \nu_{\mathrm{trap}}\left(\epsilon\right)\right\rangle \right)\left\langle \psi\right\rangle _{\mathrm{detrap}}\nonumber \\
 &  & -n\left\langle \psi\nu_{\mathrm{loss}}^{\left(\mathrm{free}\right)}\left(\epsilon\right)\right\rangle ,\label{eq:generalBalance}
\end{eqnarray}
where the velocity average $\left\langle \cdot\right\rangle $ is
defined by Eq. \eqref{eq:velocityAverage}, while $\left\langle \cdot\right\rangle _{\mathrm{coll}}$
and $\left\langle \cdot\right\rangle _{\mathrm{detrap}}$ are defined
as
\begin{align}
\left\langle \psi\left(\mathbf{v}\right)\right\rangle _{\mathrm{coll}} & \equiv\int\mathrm{d}\mathbf{v}\psi\left(\mathbf{v}\right)\tilde{w}_{\mathrm{coll}}\left(v\right),\\
\left\langle \psi\left(\mathbf{v}\right)\right\rangle _{\mathrm{detrap}} & \equiv\int\mathrm{d}\mathbf{v}\psi\left(\mathbf{v}\right)\tilde{w}_{\mathrm{detrap}}\left(v\right).
\end{align}
By choosing $\psi\left(\mathbf{v}\right)=1$, $\psi\left(\mathbf{v}\right)=m\mathbf{v}$
and $\psi\left(\mathbf{v}\right)=\epsilon\equiv\frac{1}{2}mv^{2}$,
respective balance equations for free particle continuity, momentum
and energy result:
\begin{eqnarray}
\frac{\partial}{\partial t}n+\frac{\partial}{\partial\mathbf{r}}\cdot n\left\langle \mathbf{v}\right\rangle  & = & -n\left\langle \nu_{\mathrm{trap}}\left(\epsilon\right)\right\rangle +\Phi\left(t\right)\ast\left(n\left\langle \nu_{\mathrm{trap}}\left(\epsilon\right)\right\rangle \right)\nonumber \\
 &  & -n\left\langle \nu_{\mathrm{loss}}^{\left(\mathrm{free}\right)}\left(\epsilon\right)\right\rangle ,\label{eq:continuityBalance}\\
\frac{\partial}{\partial t}n\left\langle m\mathbf{v}\right\rangle +\frac{\partial}{\partial\mathbf{r}}\cdot n\left\langle m\mathbf{v}\mathbf{v}\right\rangle -e\mathbf{E}n & = & -n\left\langle m\mathbf{v}\nu_{\mathrm{coll}}\left(\epsilon\right)\right\rangle \nonumber \\
 &  & -n\left\langle m\mathbf{v}\nu_{\mathrm{trap}}\left(\epsilon\right)\right\rangle \nonumber \\
 &  & -n\left\langle m\mathbf{v}\nu_{\mathrm{loss}}^{\left(\mathrm{free}\right)}\left(\epsilon\right)\right\rangle ,\label{eq:momentumBalance}\\
\frac{\partial}{\partial t}n\left\langle \epsilon\right\rangle +\frac{\partial}{\partial\mathbf{r}}\cdot n\left\langle \epsilon\mathbf{v}\right\rangle -e\mathbf{E}\cdot n\left\langle \mathbf{v}\right\rangle  & = & -n\left\langle \epsilon\nu_{\mathrm{coll}}\left(\epsilon\right)\right\rangle +n\left\langle \nu_{\mathrm{coll}}\left(\epsilon\right)\right\rangle \left\langle \epsilon\right\rangle _{\mathrm{coll}}\nonumber \\
 &  & -n\left\langle \epsilon\nu_{\mathrm{trap}}\left(\epsilon\right)\right\rangle +\Phi\left(t\right)\ast\left(n\left\langle \nu_{\mathrm{trap}}\left(\epsilon\right)\right\rangle \right)\left\langle \epsilon\right\rangle _{\mathrm{detrap}}\nonumber \\
 &  & -n\left\langle \epsilon\nu_{\mathrm{loss}}^{\left(\mathrm{free}\right)}\left(\epsilon\right)\right\rangle .\label{eq:energyBalance}
\end{eqnarray}
The latter two equations can be written explicitly as differential
equations in the average momentum and energy by expanding time derivatives
and applying the continuity equation \eqref{eq:continuityBalance}:
\begin{eqnarray}
n\frac{\partial}{\partial t}\left\langle m\mathbf{v}\right\rangle +\frac{\partial}{\partial\mathbf{r}}\cdot n\left\langle m\mathbf{v}\mathbf{v}\right\rangle -\left\langle m\mathbf{v}\right\rangle \frac{\partial}{\partial\mathbf{r}}\cdot n\left\langle \mathbf{v}\right\rangle -e\mathbf{E}n & = & -n\left\langle m\mathbf{v}\nu_{\mathrm{coll}}\left(\epsilon\right)\right\rangle \nonumber \\
 &  & -n\left\langle m\mathbf{v}\nu_{\mathrm{trap}}\left(\epsilon\right)\right\rangle +n\left\langle m\mathbf{v}\right\rangle \left\langle \nu_{\mathrm{trap}}\left(\epsilon\right)\right\rangle -\left\langle m\mathbf{v}\right\rangle \Phi\left(t\right)\ast\left(n\left\langle \nu_{\mathrm{trap}}\left(\epsilon\right)\right\rangle \right)\nonumber \\
 &  & -n\left\langle m\mathbf{v}\nu_{\mathrm{loss}}^{\left(\mathrm{free}\right)}\left(\epsilon\right)\right\rangle +n\left\langle m\mathbf{v}\right\rangle \left\langle \nu_{\mathrm{loss}}^{\left(\mathrm{free}\right)}\left(\epsilon\right)\right\rangle ,\\
n\frac{\partial}{\partial t}\left\langle \epsilon\right\rangle +\frac{\partial}{\partial\mathbf{r}}\cdot n\left\langle \epsilon\mathbf{v}\right\rangle -\left\langle \epsilon\right\rangle \frac{\partial}{\partial\mathbf{r}}\cdot n\left\langle \mathbf{v}\right\rangle -e\mathbf{E}\cdot n\left\langle \mathbf{v}\right\rangle  & = & -n\left\langle \epsilon\nu_{\mathrm{coll}}\left(\epsilon\right)\right\rangle +n\left\langle \nu_{\mathrm{coll}}\left(\epsilon\right)\right\rangle \left\langle \epsilon\right\rangle _{\mathrm{coll}}\nonumber \\
 &  & -n\left\langle \epsilon\nu_{\mathrm{trap}}\left(\epsilon\right)\right\rangle +n\left\langle \epsilon\right\rangle \left\langle \nu_{\mathrm{trap}}\left(\epsilon\right)\right\rangle -\left(\left\langle \epsilon\right\rangle -\left\langle \epsilon\right\rangle _{\mathrm{detrap}}\right)\Phi\left(t\right)\ast\left(n\left\langle \nu_{\mathrm{trap}}\left(\epsilon\right)\right\rangle \right)\nonumber \\
 &  & -n\left\langle \epsilon\nu_{\mathrm{loss}}^{\left(\mathrm{free}\right)}\left(\epsilon\right)\right\rangle +n\left\langle \epsilon\right\rangle \left\langle \nu_{\mathrm{loss}}^{\left(\mathrm{free}\right)}\left(\epsilon\right)\right\rangle .
\end{eqnarray}
Solution of these balance equations requires some approximation in
the evaluation of the averages of the collision frequencies. In what
follows we solve these balance equations using momentum transfer theory
\citep{Robson1984} to develop expressions for the mobility, diffusion
and the mean energy in terms of the underlying microscopic frequencies
for collisions, trapping and losses. Application of these relationships
yield some interesting phenomenon including negative differential
conductivity (NDC) and heating/cooling, as well as conditions on the
relevant frequencies for such phenomena to occur.

\section{\label{sec:Phenomena}Mobility and the Wannier energy relation: Heating/cooling
and NDC}

In this section, we are interested in physical properties in the weak-gradient
hydrodynamic regime. In this limit, properties that are intensive
(independent of particle number) become time invariant and spatial
gradients vanish \citep{Robson1986}, resulting in simplified momentum
and energy balance equations that provide expressions for the applied
acceleration and power input by the field:
\begin{eqnarray}
\frac{e\mathbf{E}}{m} & = & \left\langle \mathbf{v}\nu_{\mathrm{coll}}\left(\epsilon\right)\right\rangle ^{\left(0\right)}\nonumber \\
 &  & +\left\langle \mathbf{v}\nu_{\mathrm{trap}}\left(\epsilon\right)\right\rangle ^{\left(0\right)}-\left(1-R\right)\mathbf{W}\left\langle \nu_{\mathrm{trap}}\left(\epsilon\right)\right\rangle ^{\left(0\right)}\nonumber \\
 &  & +\left\langle \mathbf{v}\nu_{\mathrm{loss}}^{\left(\mathrm{free}\right)}\left(\epsilon\right)\right\rangle ^{\left(0\right)}-\mathbf{W}\left\langle \nu_{\mathrm{loss}}^{\left(\mathrm{free}\right)}\left(\epsilon\right)\right\rangle ^{\left(0\right)},\\
e\mathbf{E}\cdot\mathbf{W} & = & \left\langle \epsilon\nu_{\mathrm{coll}}\left(\epsilon\right)\right\rangle ^{\left(0\right)}-\left\langle \nu_{\mathrm{coll}}\left(\epsilon\right)\right\rangle ^{\left(0\right)}\left\langle \epsilon\right\rangle _{\mathrm{coll}}\nonumber \\
 &  & +\left\langle \epsilon\nu_{\mathrm{trap}}\left(\epsilon\right)\right\rangle ^{\left(0\right)}-\varepsilon\left\langle \nu_{\mathrm{trap}}\left(\epsilon\right)\right\rangle ^{\left(0\right)}+R\left\langle \nu_{\mathrm{trap}}\left(\epsilon\right)\right\rangle ^{\left(0\right)}\left(\varepsilon-\left\langle \epsilon\right\rangle _{\mathrm{detrap}}\right)\nonumber \\
 &  & +\left\langle \epsilon\nu_{\mathrm{loss}}^{\left(\mathrm{free}\right)}\left(\epsilon\right)\right\rangle ^{\left(0\right)}-\varepsilon\left\langle \nu_{\mathrm{loss}}^{\left(\mathrm{free}\right)}\left(\epsilon\right)\right\rangle ^{\left(0\right)}.
\end{eqnarray}
where the superscript ``$\left(0\right)$'' denotes that quantities
are in the steady, spatially uniform state. Here, the moments for
drift velocity and mean energy have been respectively defined
\begin{eqnarray}
\mathbf{W} & \equiv & \left\langle \mathbf{v}\right\rangle ^{\left(0\right)},\\
\varepsilon & \equiv & \left\langle \epsilon\right\rangle ^{\left(0\right)},
\end{eqnarray}
and we have introduced the quantity $R$ as the steady-state ratio
of the number of particles leaving traps to those entering traps:
\begin{equation}
R\equiv\left(\frac{\Phi\left(t\right)\ast n\left(t,\mathbf{r}\right)}{n\left(t,\mathbf{r}\right)}\right)^{\left(0\right)}\equiv\lim_{t\rightarrow\infty}\frac{\Phi\left(t\right)\ast N\left(t\right)}{N\left(t\right)}.\label{eq:Rdefinition}
\end{equation}
In the following subsections, we make these balance equations more
useful by using momentum transfer theory to approximate the velocity
averages of the form $\left\langle \nu\left(\epsilon\right)\right\rangle $,
$\left\langle \mathbf{v}\nu\left(\epsilon\right)\right\rangle $ and
$\left\langle \epsilon\nu\left(\epsilon\right)\right\rangle $. The
simplified balance equations that result provide expressions for particle
mobility and mean energy which in turn can be used to quantify heating/cooling
and to explore NDC.

\subsection{Momentum transfer theory}

Momentum-transfer theory \citep{Robson1984} enables a systematic
procedure for evaluating the average rates detailed above. In this
procedure, process rates, $\nu\left(\epsilon\right)$, are expanded
about some representative energy, which we take to be the mean energy,
$\varepsilon$:
\begin{equation}
\nu\left(\epsilon\right)=\sum_{i\geq0}\frac{\nu^{\left(i\right)}\left(\varepsilon\right)}{i!}\left(\epsilon-\varepsilon\right)^{i},\label{eq:MTapproximation}
\end{equation}
where the superscript ``$\left(i\right)$'' denotes the $i$-th
energy derivative. This expansion can then be truncated to the desired
order of accuracy. By truncating to just the initial constant term,
we have zeroth-order momentum transfer theory, which provides a mobility
and a Wannier energy relation that is sufficient for exploring NDC
and energy-independent heating/cooling. For heating/cooling that varies
with energy, we must truncate the above expansion linearly and use
first-order momentum transfer theory.

\subsubsection{\label{subsec:BalanceZeroth}Zeroth-order momentum transfer theory}

Truncating the energy expansion, Eq. \eqref{eq:MTapproximation},
to the constant term gives the zeroth-order momentum transfer theory
approximation
\begin{equation}
\left\langle \psi\left(\mathbf{v}\right)\nu\left(\epsilon\right)\right\rangle \approx\left\langle \psi\left(\mathbf{v}\right)\right\rangle \nu\left(\varepsilon\right).\label{eq:constantEnergy}
\end{equation}
This approximation yields results that are functionally equivalent
to what arises for the case of constant process rates, as considered
in \citep{Stokes2016}, but with some functional dependence on the
representative energy $\varepsilon$. Substituting this approximation
into the momentum and energy balance equations \eqref{eq:momentumBalance}
and \eqref{eq:energyBalance} yields
\begin{eqnarray}
\frac{e\mathbf{E}}{m} & = & \mathbf{W}\nu_{\mathrm{eff}}\left(\varepsilon\right),\label{eq:momentumBalanceZeroth}\\
e\mathbf{E}\cdot\mathbf{W} & = & \left[\varepsilon-\frac{3}{2}k_{\mathrm{B}}T_{\mathrm{eff}}\left(\varepsilon\right)\right]\nu_{\mathrm{eff}}\left(\varepsilon\right),\label{eq:energyBalanceZeroth}
\end{eqnarray}
where we have introduced an effective frequency
\begin{equation}
\nu_{\mathrm{eff}}\left(\varepsilon\right)\equiv\nu_{\mathrm{coll}}\left(\varepsilon\right)+R\nu_{\mathrm{trap}}\left(\varepsilon\right),\label{eq:effectiveFrequency}
\end{equation}
and an energy-dependent effective temperature, written as a weighted
sum of the two Maxwellian source temperatures
\begin{equation}
T_{\mathrm{eff}}\left(\varepsilon\right)\equiv\omega_{\mathrm{coll}}\left(\varepsilon\right)T_{\mathrm{coll}}+\omega_{\mathrm{detrap}}\left(\varepsilon\right)T_{\mathrm{detrap}},
\end{equation}
with energy-dependent weights defined
\begin{eqnarray}
\omega_{\mathrm{coll}}\left(\varepsilon\right) & \equiv & \frac{\nu_{\mathrm{coll}}\left(\varepsilon\right)}{\nu_{\mathrm{coll}}\left(\varepsilon\right)+R\nu_{\mathrm{trap}}\left(\varepsilon\right)},\label{eq:collWeight}\\
\omega_{\mathrm{trap}}\left(\varepsilon\right) & \equiv & \frac{R\nu_{\mathrm{trap}}\left(\varepsilon\right)}{\nu_{\mathrm{coll}}\left(\varepsilon\right)+R\nu_{\mathrm{trap}}\left(\varepsilon\right)}.\label{eq:trapWeight}
\end{eqnarray}
It should be noted that, as free particle recombination and trapping
rates are constant here, the limit definition of $R$ in Eq. \eqref{eq:Rdefinition}
can be evaluated to provide the alternative implicit definition \citep{Stokes2016}:
\begin{equation}
R\equiv\int_{0}^{\infty}\mathrm{d}t\,\Phi\left(t\right)\mathrm{e}^{\left[\nu_{\mathrm{loss}}^{\left(\mathrm{free}\right)}\left(\varepsilon\right)+\nu_{\mathrm{trap}}\left(\varepsilon\right)\left(1-R\right)\right]t}.\label{eq:Rimplicit}
\end{equation}
This implicit definition can be solved analytically for $R$ only
for certain choices of the effective waiting time distribution $\Phi\left(t\right)$.
A table of such $R$ values for a variety of corresponding $\Phi\left(t\right)$
is presented in Appendix A of \citep{Stokes2016}.

The zeroth-order momentum balance equation \eqref{eq:momentumBalanceZeroth}
provides the drift velocity in terms of the electric field $\mathbf{E}$:
\begin{equation}
\mathbf{W}\equiv K\mathbf{E},\label{eq:driftVelocity}
\end{equation}
where the constant of proportionality $K$ defines the charged particle
mobility
\begin{equation}
K\equiv\frac{e}{m\nu_{\mathrm{eff}}\left(\varepsilon\right)}.\label{eq:zerothMobility}
\end{equation}
We observe that the mobility is inversely proportional to both collision
and trapping process rates through the effective frequency defined
in Eq. \eqref{eq:effectiveFrequency}. This result is expected as
both the scattering and detrapping processes occur isotropically.
Evidently, precisely how mobility varies with energy depends entirely
on the energy dependence of the process frequencies.

Using both the momentum and energy balance equations \eqref{eq:momentumBalanceZeroth}
and \eqref{eq:energyBalanceZeroth}, we can also find the Wannier
energy relation for the average energy 
\begin{equation}
\varepsilon=\frac{3}{2}k_{\mathrm{B}}T_{\mathrm{eff}}\left(\varepsilon\right)+mW^{2}.\label{eq:zerothWannier}
\end{equation}
We can confirm that when there is no trapping, $\nu_{\mathrm{trap}}\left(\varepsilon\right)=0$,
the mobility and Wannier energy relation reduce to the classical results
valid for dilute gaseous systems \citep{Robson1986}:
\begin{eqnarray}
K & = & \frac{e}{m\nu_{\mathrm{coll}}\left(\varepsilon\right)},\\
\varepsilon & = & \frac{3}{2}k_{\mathrm{B}}T_{\mathrm{coll}}+mW^{2}.
\end{eqnarray}
The zeroth-order mobility and Wannier energy relation derived here
are used to describe energy-independent heating/cooling in Secs. \ref{sec:HotCold-1}
and \ref{sec:HotCold-2} as well as NDC in Sec. \ref{sec:NDC}.

\subsubsection{\label{sec:BalanceFirst}First-order momentum transfer theory}

Including an additional term in the energy expansion, Eq. \eqref{eq:MTapproximation},
gives the first-order momentum transfer theory approximation
\begin{equation}
\left\langle \psi\left(\mathbf{v}\right)\nu\left(\epsilon\right)\right\rangle \approx\left\langle \psi\left(\mathbf{v}\right)\right\rangle \nu\left(\varepsilon\right)+\left\langle \psi\left(\mathbf{v}\right)\left(\epsilon-\varepsilon\right)\right\rangle \nu^{\prime}\left(\varepsilon\right),\label{eq:linearEnergy}
\end{equation}
where $\nu^{\prime}\left(\varepsilon\right)$ denotes the energy derivative
of $\nu\left(\varepsilon\right)$. Substitution into the momentum
and energy balance equations \eqref{eq:momentumBalance} and \eqref{eq:energyBalance}
yields
\begin{eqnarray}
\frac{e\mathbf{E}}{m} & = & \mathbf{W}\nu_{\mathrm{eff}}\left(\varepsilon\right)+\mathrm{cov}\left(\mathbf{v},\epsilon\right)\nu_{\mathrm{total}}^{\prime}\left(\varepsilon\right),\label{eq:momentumBalanceFirst}\\
e\mathbf{E}\cdot\mathbf{W} & = & \left[\varepsilon-\frac{3}{2}k_{\mathrm{B}}T_{\mathrm{eff}}\left(\varepsilon\right)\right]\nu_{\mathrm{eff}}\left(\varepsilon\right)+\mathrm{var}\left(\epsilon\right)\nu_{\mathrm{total}}^{\prime}\left(\varepsilon\right)\nonumber \\
 &  & -\frac{\frac{3}{2}\left(k_{\mathrm{B}}T_{\mathrm{coll}}\right)^{2}\nu_{\mathrm{coll}}^{\prime}\left(\varepsilon\right)}{1+\left(\frac{3}{2}k_{\mathrm{B}}T_{\mathrm{coll}}-\varepsilon\right)\frac{\nu_{\mathrm{coll}}^{\prime}\left(\varepsilon\right)}{\nu_{\mathrm{coll}}\left(\varepsilon\right)}}-\frac{\frac{3}{2}\left(k_{\mathrm{B}}T_{\mathrm{detrap}}\right)^{2}R\nu_{\mathrm{trap}}^{\prime}\left(\varepsilon\right)}{1+\left(\frac{3}{2}k_{\mathrm{B}}T_{\mathrm{detrap}}-\varepsilon\right)\frac{\nu_{\mathrm{trap}}^{\prime}\left(\varepsilon\right)}{\nu_{\mathrm{trap}}\left(\varepsilon\right)}},\label{eq:energyBalanceFirst}
\end{eqnarray}
where we define $\nu_{\mathrm{total}}\left(\varepsilon\right)\equiv\nu_{\mathrm{coll}}\left(\varepsilon\right)+\nu_{\mathrm{trap}}\left(\varepsilon\right)+\nu_{\mathrm{loss}}^{\left(\mathrm{free}\right)}\left(\varepsilon\right)$,
and higher order velocity moments have been introduced in the form
of the velocity-energy covariance
\begin{equation}
\mathrm{cov}\left(\mathbf{v},\epsilon\right)\equiv\left\langle \left(\mathbf{v}-\mathbf{W}\right)\left(\epsilon-\varepsilon\right)\right\rangle ^{\left(0\right)}\equiv\boldsymbol{\xi}-\varepsilon\mathbf{W},
\end{equation}
where $\boldsymbol{\xi}\equiv\left\langle \epsilon\mathbf{v}\right\rangle ^{\left(0\right)}$
is the energy flux, and the energy variance
\begin{equation}
\mathrm{var}\left(\epsilon\right)\equiv\left\langle \left(\epsilon-\varepsilon\right)^{2}\right\rangle ^{\left(0\right)}\equiv\left\langle \epsilon^{2}\right\rangle ^{\left(0\right)}-\varepsilon^{2}.
\end{equation}
These higher order velocity moments can be approximated using zeroth-order
momentum transfer theory, as is done in Appendix \ref{sec:HigherMoments},
to yield approximations expressed solely in terms of the lower order
velocity moments $\mathbf{W}$ and $\varepsilon$. For example, the
velocity-energy covariance can be approximated with
\begin{equation}
\mathrm{cov}\left(\mathbf{v},\epsilon\right)\approx\frac{2}{3}\left(\varepsilon+2mW^{2}\right)\mathbf{W}.
\end{equation}
Using this approximation in conjunction with the first-order momentum
balance equation \eqref{eq:momentumBalanceFirst}, we find the mobility,
as defined by Eq. \eqref{eq:driftVelocity}:
\begin{equation}
K\approx\frac{e}{m\left[\nu_{\mathrm{eff}}\left(\varepsilon\right)+\frac{2}{3}\left(\varepsilon+2mW^{2}\right)\nu_{\mathrm{total}}^{\prime}\left(\varepsilon\right)\right]}.\label{eq:firstMobility}
\end{equation}
This is of the same functional form as the zeroth-order mobility,
Eq. \eqref{eq:zerothMobility}, but with a modification to the effective
frequency in the denominator. Note that the mobility now depends explicitly
on the drift velocity, through the $2mW^{2}$ term. Terms such as
this are sometimes omitted in the literature as their contribution
is minimal when light particles are being considered \citep{Robson1986}.

As for zeroth-order momentum transfer theory, a Wannier energy relation
can be formed by combining both momentum and energy balance equations
\eqref{eq:momentumBalanceFirst} and \eqref{eq:energyBalanceFirst}:
\begin{align}
\varepsilon & =\frac{3}{2}k_{\mathrm{B}}T_{\mathrm{eff}}\left(\varepsilon\right)+mW^{2}\nonumber \\
 & -\frac{\nu_{\mathrm{total}}^{\prime}\left(\varepsilon\right)}{\nu_{\mathrm{eff}}\left(\varepsilon\right)}\mathrm{cov}\left(\epsilon,\epsilon-m\mathbf{W}\cdot\mathbf{v}\right)+\frac{\frac{3}{2}\left(k_{\mathrm{B}}T_{\mathrm{coll}}\right)^{2}\frac{\nu_{\mathrm{coll}}^{\prime}\left(\varepsilon\right)}{\nu_{\mathrm{eff}}\left(\varepsilon\right)}}{1+\left(\frac{3}{2}k_{\mathrm{B}}T_{\mathrm{coll}}-\varepsilon\right)\frac{\nu_{\mathrm{coll}}^{\prime}\left(\varepsilon\right)}{\nu_{\mathrm{coll}}\left(\varepsilon\right)}}+\frac{\frac{3}{2}\left(k_{\mathrm{B}}T_{\mathrm{detrap}}\right)^{2}\frac{R\nu_{\mathrm{trap}}^{\prime}\left(\varepsilon\right)}{\nu_{\mathrm{eff}}\left(\varepsilon\right)}}{1+\left(\frac{3}{2}k_{\mathrm{B}}T_{\mathrm{detrap}}-\varepsilon\right)\frac{\nu_{\mathrm{trap}}^{\prime}\left(\varepsilon\right)}{\nu_{\mathrm{trap}}\left(\varepsilon\right)}}.\label{eq:firstWannier}
\end{align}
This first-order Wannier energy relation is written in terms of higher
order velocity moments via the covariance
\begin{equation}
\mathrm{cov}\left(\epsilon,\epsilon-m\mathbf{W}\cdot\mathbf{v}\right)\equiv\mathrm{var}\left(\epsilon\right)-m\mathbf{W}\cdot\mathrm{cov}\left(\mathbf{v},\epsilon\right).
\end{equation}
As before, the results in Appendix \ref{sec:HigherMoments} allow
for this covariance to also be written approximately in terms of lower
order velocity moments:
\begin{equation}
\mathrm{cov}\left(\epsilon,\epsilon-m\mathbf{W}\cdot\mathbf{v}\right)\approx\frac{2}{3}\left(\varepsilon-\frac{1}{2}mW^{2}\right)^{2}+\frac{17}{6}\left(mW^{2}\right)^{2}+\frac{5}{3}\omega_{\mathrm{coll}}\left(\varepsilon\right)\omega_{\mathrm{trap}}\left(\varepsilon\right)\left[\frac{3}{2}k_{\mathrm{B}}\left(T_{\mathrm{coll}}-T_{\mathrm{detrap}}\right)\right]^{2}.
\end{equation}
This expression can be used to write the first-order Wannier energy
relation \eqref{eq:firstWannier} in an approximate closed form, independent
of higher order velocity moments.

Comparing the above first-order momentum transfer theory results for
mobility and average energy, Eqs. \eqref{eq:firstMobility} and \eqref{eq:firstWannier},
to their zeroth-order counterparts, Eqs. \eqref{eq:zerothMobility}
and \eqref{eq:zerothWannier}, provides an estimate of the error incurred
by the zeroth-order momentum transfer theory approximation.

In Sec. \ref{sec:HotCold-3}, we use the first-order mobility and
Wannier energy relation derived here to describe heating/cooling that
is due to the energy dependence of physical processes.

\subsection{\label{sec:HotCold}Heating and cooling}

In this subsection, we determine the effect that each of the physical
processes described by the generalised Boltzmann equation \eqref{eq:boltzmannEquation}
have on the average particle energy. That is, whether there is an
increase or decrease in the average energy corresponding to a respective
heating or cooling of the particles as a result of collisions, trapping
or recombination.

\subsubsection{\label{sec:HotCold-1}Collisional and trap-based heating/cooling}

To consider the effect of collisions on the average energy, we consider
the case of constant process rates where the average energy is given
by the zeroth-order Wannier energy relation \eqref{eq:zerothWannier}.
For collisions that are infrequent relative to trapping, i.e. $\nu_{\mathrm{coll}}<R\nu_{\mathrm{trap}}$,
the average energy can be written approximately to first order in
$\nu_{\mathrm{coll}}/R\nu_{\mathrm{trap}}$:
\begin{equation}
\varepsilon\approx\varepsilon_{0}+2\left(\frac{3}{2}k_{\mathrm{B}}T_{\mathrm{HC}}-\varepsilon_{0}\right)\frac{\nu_{\mathrm{coll}}}{R\nu_{\mathrm{trap}}},\label{eq:collisionFirst}
\end{equation}
where the subscript ``$0$'' denotes the collisionless case, i.e.
$\nu_{\mathrm{coll}}=0$:
\begin{eqnarray}
\varepsilon_{0} & = & \frac{3}{2}k_{\mathrm{B}}T_{\mathrm{detrap}}+mW_{0}^{2},\\
\mathbf{W}_{0} & = & \frac{e\mathbf{E}}{mR\nu_{\mathrm{trap}}},
\end{eqnarray}
and $T_{\mathrm{HC}}$ is a threshold temperature which defines the
transition between collisional heating and cooling:
\begin{equation}
T_{\mathrm{HC}}\equiv\frac{T_{\mathrm{coll}}+T_{\mathrm{detrap}}}{2}.
\end{equation}
In the event that $\varepsilon_{0}=\frac{3}{2}k_{\mathrm{B}}T_{\mathrm{HC}}$,
the first order term in the expansion above vanishes and we must instead
consider the second-order approximation:
\begin{equation}
\varepsilon\approx\varepsilon_{0}+mW_{0}^{2}\left(\frac{\nu_{\mathrm{coll}}}{R\nu_{\mathrm{trap}}}\right)^{2}.\label{eq:collisionSecond}
\end{equation}
The expansions \eqref{eq:collisionFirst} and \eqref{eq:collisionSecond}
show that the introduction of collisions cause cooling only if the
initial average energy $\varepsilon_{0}$ exceeds the threshold energy
proportional to the temperature $T_{\mathrm{HC}}$:
\begin{equation}
\varepsilon_{0}>\frac{3}{2}k_{\mathrm{B}}T_{\mathrm{HC}},\label{eq:collisionCondition}
\end{equation}
with collisional heating occurring otherwise.

These conditions can also be shown to be applicable to trap-based
heating/cooling, in which case $\varepsilon_{0}$ would denote the
trap-free mean energy with $\nu_{\mathrm{trap}}=0$.

\subsubsection{\label{sec:HotCold-2}Energy-indiscriminate recombination heating/cooling}

We now explore the possibility of recombination heating/cooling by
once again considering constant process rates. It is usually expected
that constant loss rates, which act indiscriminate of energy, result
in a decrease in particle number that affects extensive properties
but leaves intensive properties, like the average energy, unchanged
\citep{Robson1986}. Although it is true that the recombination considered
here is not selective of particle energy, the separate recombination
rates for free and trapped particles means that recombination \textit{is}
selective of whether particles are trapped or not. Indeed, the average
energy can be shown to be a function of the difference in these recombination
rates, $\Delta\nu_{\mathrm{loss}}\equiv\nu_{\mathrm{loss}}^{\left(\mathrm{free}\right)}-\nu_{\mathrm{loss}}^{\left(\mathrm{trap}\right)}$,
only becoming independent when recombination acts uniformly across
all particles, i.e. $\nu_{\mathrm{loss}}^{\left(\mathrm{free}\right)}=\nu_{\mathrm{loss}}^{\left(\mathrm{trap}\right)}$.
The recombination dependence appears in the average energy through
the quantity $R$, whose definition in Eq. \eqref{eq:Rimplicit} is
rewritten here explicitly in terms of $\Delta\nu_{\mathrm{loss}}$:
\begin{equation}
R\equiv\int_{0}^{\infty}\mathrm{d}t\,\phi\left(t\right)\mathrm{e}^{\left[\Delta\nu_{\mathrm{loss}}+\nu_{\mathrm{trap}}\left(1-R\right)\right]t}.
\end{equation}
The original definition of $R$ was given by Eq. \eqref{eq:Rdefinition}
as the steady-state ratio between the number of particles leaving
and entering traps. Without recombination, this ratio is unity as
an equilibrium arises between free and trapped particles \citep{Stokes2016}.
Even with recombination, this ratio should remain at unity so long
as the number of free and trapped particles reduce equally due to
recombination, $\Delta\nu_{\mathrm{loss}}=0$.

We explore the effect of $R$ on heating/cooling by performing a small
$\Delta\nu_{\mathrm{loss}}$ expansion:
\begin{equation}
R\approx1+\frac{\Delta\nu_{\mathrm{loss}}}{\nu_{\mathrm{detrap}}+\nu_{\mathrm{trap}}},
\end{equation}
where the detrapping rate has been introduced
\begin{equation}
\nu_{\mathrm{detrap}}^{-1}\equiv\int_{0}^{\infty}\mathrm{d}t\,\phi\left(t\right)t.
\end{equation}
Proceeding to perform a small $\Delta\nu_{\mathrm{loss}}$ expansion
of the average energy, in part by using the above expansion of $R$,
gives the average energy to first order:
\begin{equation}
\varepsilon\approx\varepsilon_{0}+2\left(\frac{3}{2}k_{\mathrm{B}}T_{\mathrm{HC}}-\varepsilon_{0}\right)\frac{\nu_{\mathrm{trap}}}{\nu_{\mathrm{coll}}+\nu_{\mathrm{trap}}}\frac{\Delta\nu_{\mathrm{loss}}}{\nu_{\mathrm{detrap}}+\nu_{\mathrm{trap}}},\label{eq:recombinationFirst}
\end{equation}
where the subscript ``$0$'' denotes the case of uniform recombination,
$\Delta\nu_{\mathrm{loss}}=0$:
\begin{eqnarray}
\varepsilon_{0} & = & \frac{3}{2}k_{\mathrm{B}}T_{\mathrm{eff,0}}+mW_{0}^{2},\\
\mathbf{W}_{0} & = & \frac{e\mathbf{E}}{m\left(\nu_{\mathrm{coll}}+\nu_{\mathrm{trap}}\right)},\\
T_{\mathrm{eff,0}} & = & \frac{\nu_{\mathrm{coll}}T_{\mathrm{coll}}+\nu_{\mathrm{trap}}T_{\mathrm{detrap}}}{\nu_{\mathrm{coll}}+\nu_{\mathrm{trap}}},
\end{eqnarray}
and the threshold temperature in this case is defined as
\begin{equation}
T_{\mathrm{HC}}\equiv\frac{T_{\mathrm{eff},0}+T_{\mathrm{detrap}}}{2}.
\end{equation}
In the event that $\varepsilon_{0}=\frac{3}{2}k_{\mathrm{B}}T_{\mathrm{HC}}$,
we have instead the second-order approximation for average energy:
\begin{equation}
\varepsilon\approx\varepsilon_{0}+mW_{0}^{2}\left(\frac{\nu_{\mathrm{trap}}}{\nu_{\mathrm{coll}}+\nu_{\mathrm{trap}}}\frac{\Delta\nu_{\mathrm{loss}}}{\nu_{\mathrm{detrap}}+\nu_{\mathrm{trap}}}\right)^{2},\label{eq:recombinationSecond}
\end{equation}
From the small $\Delta\nu_{\mathrm{loss}}$ expansions \eqref{eq:recombinationFirst}
and \eqref{eq:recombinationSecond}, we see that if there is a relative
loss of free particles, $\nu_{\mathrm{loss}}^{\left(\mathrm{free}\right)}>\nu_{\mathrm{loss}}^{\left(\mathrm{trap}\right)}$,
then recombination cooling can occur if those free particles are sufficiently
energetic prior to being lost:
\begin{equation}
\varepsilon_{0}>\frac{3}{2}k_{\mathrm{B}}T_{\mathrm{HC}}.
\end{equation}
Conversely, if there is a relative gain of free particles, $\nu_{\mathrm{loss}}^{\left(\mathrm{free}\right)}<\nu_{\mathrm{loss}}^{\left(\mathrm{trap}\right)}$,
then recombination cooling can occur if those free particles are sufficiently
cold to begin with:
\begin{equation}
\varepsilon_{0}<\frac{3}{2}k_{\mathrm{B}}T_{\mathrm{HC}}.
\end{equation}
Overall, for distinct free and trapped particle recombination rates
such that $\nu_{\mathrm{loss}}^{\left(\mathrm{free}\right)}\neq\nu_{\mathrm{loss}}^{\left(\mathrm{trap}\right)}$,
the condition for recombination cooling can be summarised as
\begin{equation}
\left(\varepsilon_{0}-\frac{3}{2}k_{\mathrm{B}}T_{\mathrm{HC}}\right)\Delta\nu_{\mathrm{loss}}>0,\label{eq:recombinationCondition1}
\end{equation}
with recombination heating occurring otherwise.

\subsubsection{\label{sec:HotCold-3}Energy-selective recombination heating/cooling}

In the event that no traps are present, $\nu_{\mathrm{trap}}=0$,
or where recombination acts uniformly across all free and trapped
particles, $\nu_{\mathrm{loss}}^{\left(\mathrm{free}\right)}=\nu_{\mathrm{loss}}^{\left(\mathrm{trap}\right)}$,
heating and cooling can not occur due to the trap-selective recombination
described in the previously. In this case, heating or cooling can
only occur if recombination acts selectively based on the energy of
the free particles. To show this, we will consider the first-order
Wannier energy relation \eqref{eq:firstWannier} with constant collision
and trapping rates and constant free particle recombination rate energy
derivative $\nu_{\mathrm{loss}}^{\left(\mathrm{free}\right)\prime}$.
Performing a small $\nu_{\mathrm{loss}}^{\left(\mathrm{free}\right)\prime}/\nu_{\mathrm{eff}}$
expansion of this average energy gives, to first order:
\begin{equation}
\varepsilon\approx\varepsilon_{0}-\left\{ \frac{2}{3}\left(\varepsilon_{0}+\frac{1}{2}mW_{0}^{2}\right)^{2}+\frac{11}{2}\left(mW_{0}^{2}\right)^{2}+\frac{5}{3}\omega_{\mathrm{coll}}\omega_{\mathrm{trap}}\left[\frac{3}{2}k_{\mathrm{B}}\left(T_{\mathrm{coll}}-T_{\mathrm{detrap}}\right)\right]^{2}\right\} \frac{\nu_{\mathrm{loss}}^{\left(\mathrm{free}\right)\prime}}{\nu_{\mathrm{eff}}},\label{eq:recombinationEnergyFirst}
\end{equation}
where the subscript ``$0$'' denotes no energy dependence in the
free particle recombination rate, $\nu_{\mathrm{loss}}^{\left(\mathrm{free}\right)\prime}=0$:
\begin{eqnarray}
\varepsilon_{0} & = & \frac{3}{2}k_{\mathrm{B}}T_{\mathrm{eff}}+mW_{0}^{2},\\
\mathbf{W}_{0} & = & \frac{e\mathbf{E}}{m\nu_{\mathrm{eff}}}.
\end{eqnarray}
As is expected, the expansion \eqref{eq:recombinationEnergyFirst}
suggests that recombination cooling occurs when recombination is selective
of higher energy particles,
\begin{equation}
\nu_{\mathrm{loss}}^{\left(\mathrm{free}\right)\prime}>0,\label{eq:recombinationCondition2}
\end{equation}
with recombination heating occurring when it is selective of lower
energy particles. This confirms for this model the well known phenomena
of attachment heating/cooling \citep{Robson1986}.

\subsection{\label{sec:NDC}Negative differential conductivity}

Negative differential conductivity (NDC) occurs when an \textit{increase}
in field strength causes a \textit{decrease} in the drift velocity
\citep{Robson1984}: 
\begin{equation}
\frac{\mathrm{d}W}{\mathrm{d}E}<0.
\end{equation}
The field rate of change of drift velocity can be found directly from
the zeroth-order Wannier energy relation \eqref{eq:zerothWannier}
as
\begin{equation}
\frac{\mathrm{d}W}{\mathrm{d}E}=\frac{1}{2mW}\left[1-\frac{3}{2}k_{\mathrm{B}}T_{\mathrm{eff}}^{\prime}\left(\varepsilon\right)\right]\frac{\mathrm{d}\varepsilon}{\mathrm{d}E},
\end{equation}
which provides the condition for the occurrence of NDC:
\begin{equation}
\frac{3}{2}k_{\mathrm{B}}T_{\mathrm{eff}}^{\prime}\left(\varepsilon\right)>1.\label{eq:ndcCondition}
\end{equation}
The NDC condition assumes that the mean energy increases monotonically
with the field
\begin{equation}
\frac{\mathrm{d}\varepsilon}{\mathrm{d}E}>0.
\end{equation}
This is equivalent to restricting the effective frequency $\nu_{\mathrm{eff}}\left(\varepsilon\right)$
so as to avoid runaway and ensure that an equilibrium is reached \citep{Skullerud1979}:
\begin{equation}
\frac{\mathrm{d}}{\mathrm{d}\varepsilon}\left(\nu_{\mathrm{eff}}\left(\varepsilon\right)\sqrt{\varepsilon-\frac{3}{2}k_{\mathrm{B}}T_{\mathrm{eff}}\left(\varepsilon\right)}\right)>0.
\end{equation}
Note that the occurrence of NDC depends solely on how the effective
temperature varies with energy. This energy rate of change is proportional
to the difference in Maxwellian temperatures:
\begin{equation}
T_{\mathrm{eff}}^{\prime}\left(\varepsilon\right)=\left(T_{\mathrm{coll}}-T_{\mathrm{detrap}}\right)\omega_{\mathrm{coll}}^{\prime}\left(\varepsilon\right)=\left(T_{\mathrm{detrap}}-T_{\mathrm{coll}}\right)\omega_{\mathrm{trap}}^{\prime}\left(\varepsilon\right).
\end{equation}
Hence, in comparison with Eq. \eqref{eq:ndcCondition}, we see that
NDC here cannot occur when both scattering and detrapping sources
are of equal temperature or when the relative collision or trapping
rates, $\omega_{\mathrm{coll}}\left(\varepsilon\right)$ and $\omega_{\mathrm{trap}}\left(\varepsilon\right)$
, do not vary rapidly enough with mean energy.

\begin{figure}
\includegraphics{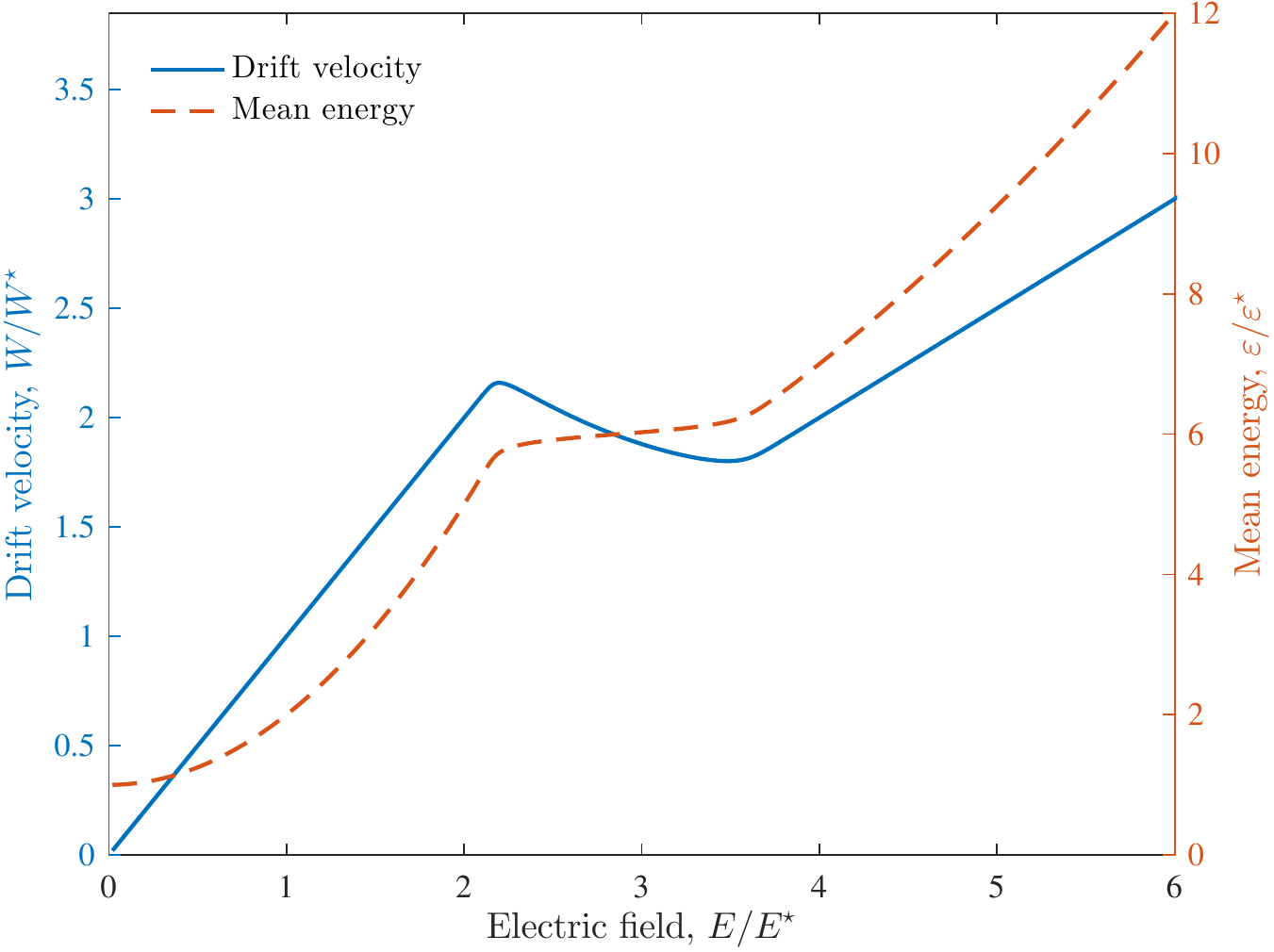}
\centering{}\caption{\label{fig:NDC}Plots of drift velocity, Eq. \eqref{eq:driftVelocity},
and mean energy, Eq. \eqref{eq:zerothWannier}, against electric field
for a situation in which negative differential conductivity arises.
All quantities have been nondimensionalised with respect to the mean
energy without a field applied, $\varepsilon^{\star}\equiv\frac{3}{2}k_{\mathrm{B}}T_{\mathrm{eff}}\left(\varepsilon^{\star}\right)$.
Specifically, we have chosen to nondimensionalise using $W^{\star}\equiv\sqrt{\frac{\varepsilon^{\star}}{m}}$
and $E^{\star}\equiv\frac{m\nu_{\mathrm{eff}}\left(\varepsilon^{\star}\right)}{e}W^{\star}$.
For this figure, we consider a constant collision frequency, $\nu_{\mathrm{coll}}\left(\varepsilon\right)=1$,
and a trapping frequency that approximates a step function, $R\nu_{\mathrm{trap}}\left(\varepsilon\right)=\frac{1}{2}\left\{ 1+\tanh\left[5\left(\varepsilon-\varepsilon_{\mathrm{thresh}}\right)\right]\right\} \approx H\left(\varepsilon-\varepsilon_{\mathrm{thresh}}\right)$,
turning on at the threshold energy $\varepsilon_{\mathrm{thresh}}=6$.
In addition, Maxwellian temperatures have been chosen such that $k_{\mathrm{B}}T_{\mathrm{coll}}=1$
and $k_{\mathrm{B}}T_{\mathrm{detrap}}=5$. }
\end{figure}

Fig. \ref{fig:NDC} plots both the drift velocity $W$ and mean energy
$\varepsilon$ as functions of the applied electric field $E$ for
a situation in which NDC arises. Previous studies \citep{Petrovic1984,Robson1984}
found that, for inelastic processes, the signature of NDC is a rapidly-increasing
mean energy. Interestingly, the opposite is true in the example considered
for our model, with the mean energy plateauing when NDC occurs. This
contrast can be understood by considering the frequency that defines
the mobility in each case. For NDC to occur, this frequency must increase
sufficiently quickly with applied field. In the referenced studies
this frequency increases over a range of energies, causing the mean
energy to increase rapidly through this range when NDC occurs. However,
in our example in Fig. \ref{fig:NDC}, the effective frequency increases
rapidly at a particular energy, causing the mean energy to plateau
at this energy during the NDC regime.

\section{\label{sec:GER}Diffusion: Generalised Einstein relations and anisotropy}

In this section, we form a generalisation of the classical Einstein
relation between diffusivity $\mathbf{D}$ and temperature $\mathbf{T}$
tensors \citep{Einstein1905}:
\begin{equation}
\frac{\mathbf{D}}{K}=\frac{k_{\mathrm{B}}\mathbf{T}}{e},\label{eq:classicalEinstein}
\end{equation}
for the phase-space model described by Eq. \eqref{eq:boltzmannEquation}.
To do this, we make use of Fick's law:
\begin{equation}
\left\langle \mathbf{v}\right\rangle \approx\mathbf{W}-\mathbf{D}\cdot\frac{1}{n}\frac{\partial n}{\partial\mathbf{r}}.\label{eq:velocityExpansion}
\end{equation}
The use of Fick's law here is justified in \citep{Stokes2016} where
it is shown that velocity averages can be written in the weak-gradient
hydrodynamic regime as a density gradient expansion
\begin{equation}
\left\langle \psi\right\rangle =\left\langle \psi\right\rangle ^{\left(0\right)}+\left\langle \psi\right\rangle ^{\left(1\right)}\cdot\frac{1}{n}\frac{\partial n}{\partial\mathbf{r}}+\left\langle \psi\right\rangle ^{\left(2\right)}\colon\frac{1}{n}\frac{\partial^{2}n}{\partial\mathbf{r}\partial\mathbf{r}}+\cdots.
\end{equation}
To find an expression for the diffusion coefficient, we must apply
density gradient expansions to all average quantities in the momentum
and energy balance equations \eqref{eq:momentumBalance} and \eqref{eq:energyBalance}.
For the mean energy we have, to first spatial order \citep{Stokes2016}
\begin{equation}
\left\langle \epsilon\right\rangle \approx\varepsilon+\boldsymbol{\gamma}\cdot\frac{1}{n}\frac{\partial n}{\partial\mathbf{r}},\label{eq:energyExpansion}
\end{equation}
where $\boldsymbol{\gamma}$ is the energy gradient parameter. Using
the density gradient expansions of average velocity and energy, Eqs.
\eqref{eq:velocityExpansion} and \eqref{eq:energyExpansion}, we
can determine the following density gradient expansions valid for
an arbitrary frequency $\nu\left(\epsilon\right)$:
\begin{eqnarray}
\left\langle \nu\left(\epsilon\right)\right\rangle  & \approx & \nu\left(\varepsilon\right)+\nu^{\prime}\left(\varepsilon\right)\boldsymbol{\gamma}\cdot\frac{1}{n}\frac{\partial n}{\partial\mathbf{r}},\\
\left\langle \mathbf{v}\nu\left(\epsilon\right)\right\rangle  & \approx & \mathbf{W}\nu\left(\varepsilon\right)+\left[\nu^{\prime}\left(\varepsilon\right)\boldsymbol{\gamma}\mathbf{W}-\nu\left(\varepsilon\right)\mathbf{D}\right]\cdot\frac{1}{n}\frac{\partial n}{\partial\mathbf{r}},\\
\left\langle \epsilon\nu\left(\epsilon\right)\right\rangle  & \approx & \varepsilon\nu\left(\varepsilon\right)+\left[\nu\left(\varepsilon\right)+\varepsilon\nu^{\prime}\left(\varepsilon\right)\right]\boldsymbol{\gamma}\cdot\frac{1}{n}\frac{\partial n}{\partial\mathbf{r}}.
\end{eqnarray}
Lastly, we also perform the density gradient expansion of the concentration
of particles leaving traps
\begin{equation}
\Phi\left(t\right)\ast n\left(t,\mathbf{r}\right)\approx Rn+\mathbf{R}^{\left(1\right)}\cdot\frac{\partial n}{\partial\mathbf{r}},
\end{equation}
where $R$ is defined by Eq. \eqref{eq:Rimplicit} as the steady-state
ratio between the number of particles leaving and entering traps,
and $\mathbf{R}^{\left(1\right)}$ is a vector that has a component
due to the energy dependence of $R$ and an intrinsic component present
even for constant process rates, as was found in Eq. (71) of \citep{Stokes2016}:
\begin{eqnarray}
\mathbf{R}^{\left(1\right)} & \equiv & R^{\prime}\left(\varepsilon\right)\boldsymbol{\gamma}+\frac{R\tau}{1+\nu_{\mathrm{trap}}\left(\varepsilon\right)R\tau}\mathbf{W},
\end{eqnarray}
where we define an average time
\begin{equation}
\tau\equiv\frac{1}{R}\int_{0}^{\infty}\mathrm{d}t\,\Phi\left(t\right)\mathrm{e}^{\left[\nu_{\mathrm{loss}}^{\left(\mathrm{free}\right)}\left(\varepsilon\right)+\nu_{\mathrm{trap}}\left(\varepsilon\right)\left(1-R\right)\right]t}t,\label{eq:timeAverage}
\end{equation}
which coincides with the mean trapping time when the free and trapped
particle recombination rates coincide, $\nu_{\mathrm{loss}}^{\left(\mathrm{free}\right)}\left(\varepsilon\right)=\nu_{\mathrm{loss}}^{\left(\mathrm{trap}\right)}$.

The weak-gradient hydrodynamic regime balance equations can now be
considered to first spatial order by applying all of the above density
gradient expansions. Doing so and equating first-order terms yields
\begin{eqnarray}
\frac{k_{\mathrm{B}}\mathbf{T}}{m} & = & \nu_{\mathrm{eff}}\left(\varepsilon\right)\mathbf{D}-\nu_{\mathrm{eff}}^{\prime}\left(\varepsilon\right)\boldsymbol{\gamma}\mathbf{W}-\frac{\nu_{\mathrm{trap}}\left(\varepsilon\right)R\tau}{1+\nu_{\mathrm{trap}}\left(\varepsilon\right)R\tau}\mathbf{W}\mathbf{W},\\
-\frac{\mathbf{Q}}{\nu_{\mathrm{eff}}\left(\varepsilon\right)} & = & \left[1-\frac{3}{2}k_{\mathrm{B}}T_{\mathrm{eff}}^{\prime}\left(\varepsilon\right)\right]\boldsymbol{\gamma}+2m\mathbf{W}\cdot\mathbf{D}\nonumber \\
 &  & +\frac{3}{2}k_{\mathrm{B}}\left(T_{\mathrm{coll}}-T_{\mathrm{detrap}}\right)\omega_{\mathrm{coll}}\left(\varepsilon\right)\omega_{\mathrm{detrap}}\left(\varepsilon\right)\frac{R\tau}{1+\nu_{\mathrm{trap}}\left(\varepsilon\right)R\tau}\mathbf{W},
\end{eqnarray}
where the temperature $\mathbf{T}$ and heat flux $\mathbf{Q}$ are
defined in terms of the peculiar velocity $\mathbf{V}\equiv\mathbf{v}-\mathbf{W}$
as
\begin{eqnarray}
k_{\mathrm{B}}\mathbf{T} & \equiv & m\left\langle \mathbf{V}\mathbf{V}\right\rangle ^{\left(0\right)},\\
\mathbf{Q} & \equiv & \frac{1}{2}m\left\langle V^{2}\mathbf{V}\right\rangle ^{\left(0\right)}.
\end{eqnarray}
By writing the above system of equations in terms of components of
diffusivity and temperature perpendicular and parallel to the field:
\begin{eqnarray}
\mathbf{D} & \equiv & D_{\perp}\left(\mathbf{I}-\hat{\mathbf{E}}\hat{\mathbf{E}}\right)+D_{\parallel}\hat{\mathbf{E}}\hat{\mathbf{E}},\\
\mathbf{T} & \equiv & T_{\perp}\left(\mathbf{I}-\hat{\mathbf{E}}\hat{\mathbf{E}}\right)+T_{\parallel}\hat{\mathbf{E}}\hat{\mathbf{E}},
\end{eqnarray}
and solving for each component of diffusivity separately yields the
generalised Einstein relations
\begin{eqnarray}
D_{\perp} & = & \frac{k_{\mathrm{B}}T_{\perp}}{m\nu_{\mathrm{eff}}\left(\varepsilon\right)},\\
D_{\parallel} & = & \frac{k_{\mathrm{B}}T_{\parallel}+mW^{2}\frac{\nu_{\mathrm{trap}}\left(\varepsilon\right)R\tau}{1+\nu_{\mathrm{trap}}\left(\varepsilon\right)R\tau}-\left[\frac{Q}{W}+\frac{3}{2}k_{\mathrm{B}}\left(T_{\mathrm{coll}}-T_{\mathrm{detrap}}\right)\frac{\nu_{\mathrm{coll}}\left(\varepsilon\right)}{\nu_{\mathrm{eff}}\left(\varepsilon\right)}\frac{\nu_{\mathrm{trap}}\left(\varepsilon\right)R\tau}{1+\nu_{\mathrm{trap}}\left(\varepsilon\right)R\tau}\right]\frac{mW^{2}}{1-\frac{3}{2}k_{\mathrm{B}}T_{\mathrm{eff}}^{\prime}\left(\varepsilon\right)}\frac{\nu_{\mathrm{eff}}^{\prime}\left(\varepsilon\right)}{\nu_{\mathrm{eff}}\left(\varepsilon\right)}}{m\nu_{\mathrm{eff}}\left(\varepsilon\right)\left(1+\frac{2mW^{2}}{1-\frac{3}{2}k_{\mathrm{B}}T_{\mathrm{eff}}^{\prime}\left(\varepsilon\right)}\frac{\nu_{\mathrm{eff}}^{\prime}\left(\varepsilon\right)}{\nu_{\mathrm{eff}}\left(\varepsilon\right)}\right)}.
\end{eqnarray}
Using the zeroth-order mobility and Wannier energy relation derived
in Sec. \ref{subsec:BalanceZeroth}, we find the identity:
\begin{equation}
\frac{\frac{\mathrm{d}\ln K}{\mathrm{d}\ln E}}{1+\frac{\mathrm{d}\ln K}{\mathrm{d}\ln E}}\equiv-\frac{2mW^{2}}{1-\frac{3}{2}k_{\mathrm{B}}T_{\mathrm{eff}}^{\prime}\left(\varepsilon\right)}\frac{\nu_{\mathrm{eff}}^{\prime}\left(\varepsilon\right)}{\nu_{\mathrm{eff}}\left(\varepsilon\right)},
\end{equation}
which allows the above generalised Einstein relations to be written
in terms of the field-dependence of the mobility $K$:
\begin{eqnarray}
\frac{D_{\perp}}{K} & = & \frac{k_{\mathrm{B}}T_{\perp}}{e},\label{eq:generalisedEinsteinPerp}\\
\frac{D_{\parallel}}{K} & = & \frac{k_{\mathrm{B}}T_{\parallel}+mW^{2}\frac{\nu_{\mathrm{trap}}\left(\varepsilon\right)R\tau}{1+\nu_{\mathrm{trap}}\left(\varepsilon\right)R\tau}}{e}\left[1+\left(1+\Delta\right)\frac{\mathrm{d}\ln K}{\mathrm{d}\ln E}\right],\label{eq:generalisedEinsteinPara}
\end{eqnarray}
where
\begin{equation}
\Delta\equiv\frac{Q+\frac{3}{2}k_{\mathrm{B}}\left(T_{\mathrm{coll}}-T_{\mathrm{detrap}}\right)W\frac{\nu_{\mathrm{coll}}\left(\varepsilon\right)}{\nu_{\mathrm{eff}}\left(\varepsilon\right)}\frac{\nu_{\mathrm{trap}}\left(\varepsilon\right)R\tau}{1+\nu_{\mathrm{trap}}\left(\varepsilon\right)R\tau}}{2k_{\mathrm{B}}T_{\parallel}W+2mW^{3}\frac{\nu_{\mathrm{trap}}\left(\varepsilon\right)R\tau}{1+\nu_{\mathrm{trap}}\left(\varepsilon\right)R\tau}}.
\end{equation}
We can see that the perpendicular generalised Einstein relation coincides
with the classical Einstein relation \eqref{eq:classicalEinstein}
and that the parallel one deviates from it, highlighting the anisotropic
nature of diffusion. In the case where there is no trapping, $\nu_{\mathrm{trap}}\left(\varepsilon\right)=0$,
the above parallel Einstein relation reduces to
\begin{eqnarray}
\frac{D_{\parallel}}{K} & = & \frac{k_{\mathrm{B}}T_{\parallel}}{e}\left[1+\left(1+\Delta\right)\frac{\mathrm{d}\ln K}{\mathrm{d}\ln E}\right],\label{eq:collisionEinstein}
\end{eqnarray}
with
\begin{equation}
\Delta\equiv\frac{Q}{2k_{\mathrm{B}}T_{\parallel}W},
\end{equation}
which coincides with the well-known gas-phase results \citep{Robson1976,Robson1984}.
The deviation of this collision-only generalised Einstein relation
\eqref{eq:collisionEinstein} from the classical Einstein relation
\eqref{eq:classicalEinstein} is due entirely to the energy dependence
of the process rates. Interestingly, this is not the case when trapping
is considered, as choosing constant process rates for the generalised
Einstein relation \eqref{eq:generalisedEinsteinPara} results in a
parallel diffusion coefficient that still has some enhancement:
\begin{equation}
\frac{D_{\parallel}}{K}=\frac{k_{\mathrm{B}}T_{\parallel}+mW^{2}\frac{\nu_{\mathrm{trap}}R\tau}{1+\nu_{\mathrm{trap}}R\tau}}{e}.
\end{equation}
This anisotropy is to be expected as, rather than moving with the
applied field, some particles become localised in traps only to detrap
later to contribute to the spread of free particles.

\section{\label{sec:Fractional}Consequences of fractional transport}

In our previous works \citep{Philippa2014,Stokes2016} it was shown
that, for certain choices of the trapping time distribution $\phi\left(t\right)$,
the phase-space model defined in Sec. \ref{sec:Model} can be described
by a diffusion equation with a time derivative of non-integer order.
Specifically, given an effective trapping time distribution with a
heavy tail of the form
\begin{equation}
\Phi\left(t\right)\sim t^{-\left(1+\alpha\right)},\label{eq:heavyTail}
\end{equation}
where $0<\alpha<1$, the phase-space model \eqref{eq:boltzmannEquation}
can be described by a Caputo time-fractional diffusion equation of
order $\alpha$ \citep{Stokes2016}. Here, the quantity $\alpha$
describes how severe traps are, with smaller values of $\alpha$ corresponding
to longer-lived traps. Long-lived traps, as described by trapping
time distributions of the form of Eq. \eqref{eq:heavyTail}, are necessary
for fractional transport to occur. Indeed, such heavy-tailed distributions
have a mean trapping time that diverges:
\begin{equation}
\int_{0}^{\infty}\mathrm{d}t\,\Phi\left(t\right)t\longrightarrow\infty.\label{eq:meanTime}
\end{equation}
However, it should be noted that to ensure transport is fractional,
there must be no trap-based recombination, $\nu_{\mathrm{loss}}^{\left(\mathrm{trap}\right)}=0$,
as such losses would cause trapped states to end prematurely and cause
the above mean trapping time to converge.

In this section, we explore consequences of fractional transport on
the results derived in the earlier sections.

\subsection{Time-of-flight current transients for fractional transport}

Plotting the current in a time-of-flight experiment versus time takes
on a signature form when transport is dispersive. That is, two power-law
regimes arise whose exponents sum to $-2$. Specifically, for a trapping
time distribution of the asymptotic form of Eq. \eqref{eq:heavyTail},
these exponents are $-\left(1-\alpha\right)$ and $-\left(1+\alpha\right)$
\citep{Scher1975}. This signature has been observed experimentally
in a variety of physical systems, including charge-carrier transport
in amorphous semiconductors \citep{Scher1975,scher1991time} and electron
transport in liquid neon \citep{Sakai1992}.

\begin{figure}
\includegraphics{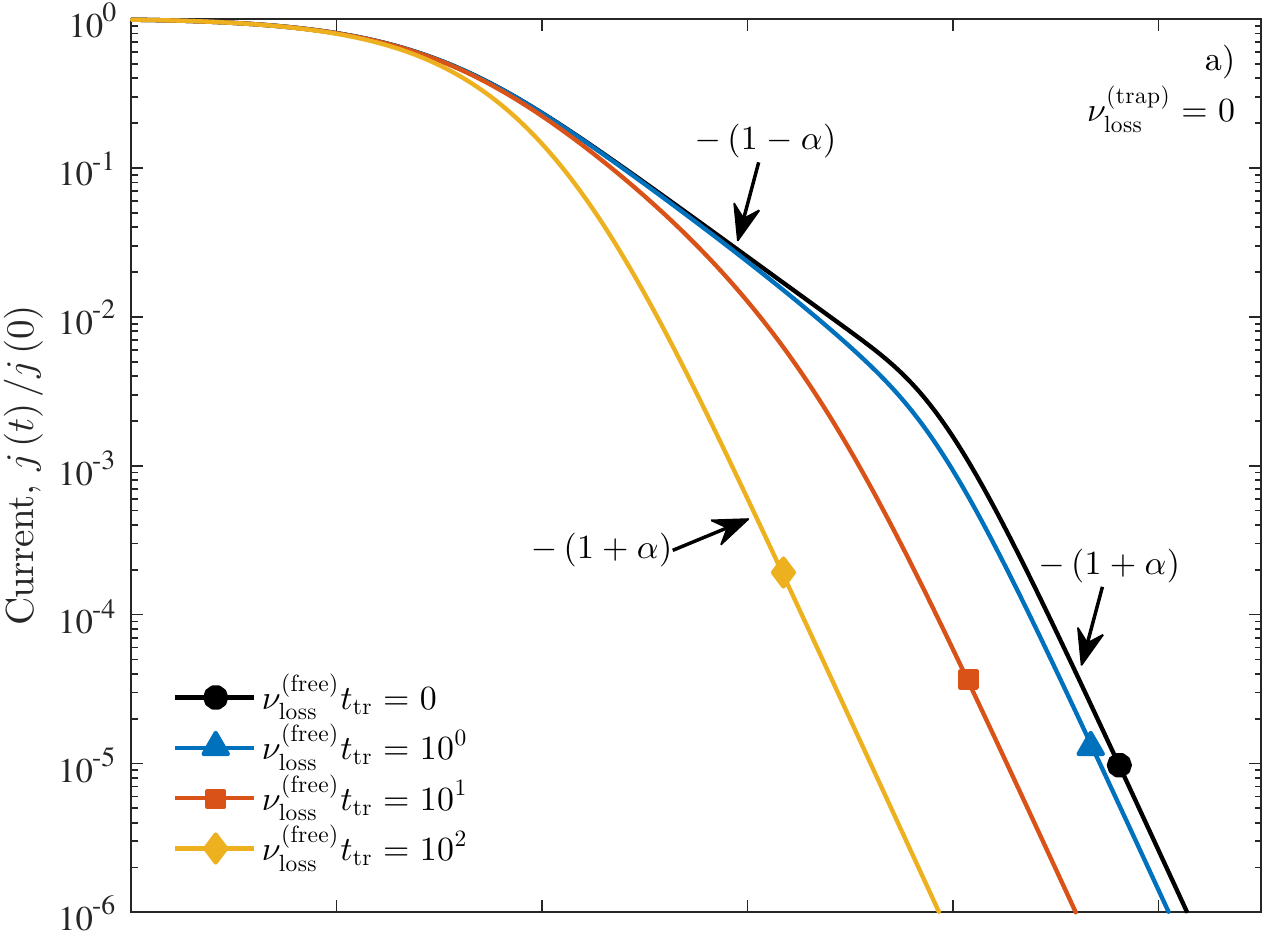}

\includegraphics{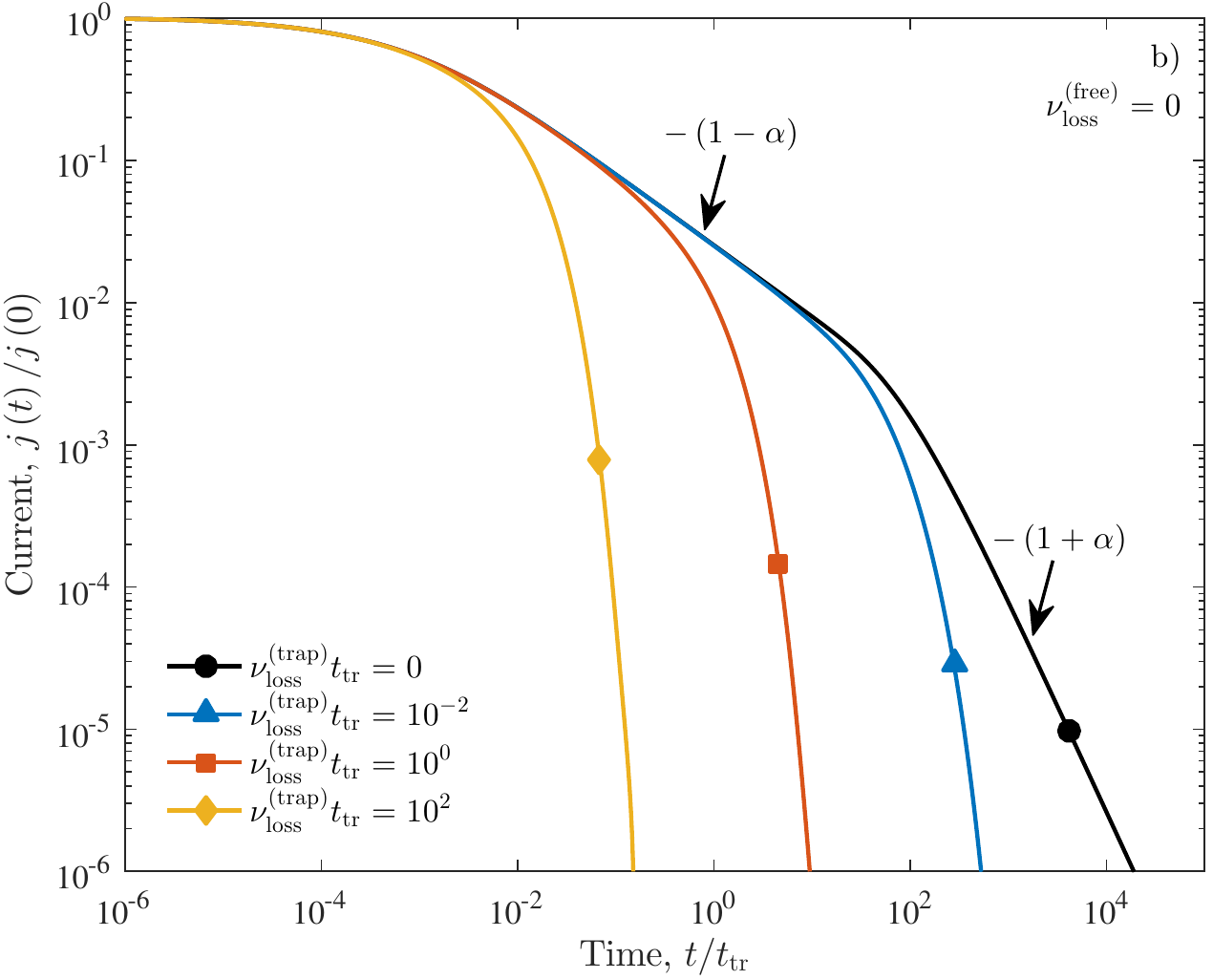}

\caption{\label{fig:currentFractional}The impact of free and trapped particle
recombination on current transients for an ideal time-of-flight experiment
as modelled by Eq. \eqref{eq:tofCurrent} for the case of dispersive
transport. Nondimensionalisation has been performed using the material
thickness $L$, trap-free transit time, $t_{\mathrm{tr}}\equiv L/W$,
and the initial current $j\left(0\right)=eN\left(0\right)/t_{\mathrm{tr}}$.
For these plots we define the diffusion coefficient, $Dt_{\mathrm{tr}}/L^{2}=0.02$,
the initial impulse location, $x_{0}/L=1/3$, and the trapping rate,
$\nu_{\mathrm{trap}}t_{\mathrm{tr}}=10^{4}$. For dispersive transport
to occur we have chosen to describe trapping times by the heavy-tailed
distribution \eqref{eq:multipleTrapping} with a trap severity of
$\alpha=1/2$. This corresponds specifically to the distribution $\phi\left(t\right)=\frac{1}{2t}\left(\frac{\sqrt{\pi}}{2}\frac{\mathrm{erf}\sqrt{\nu_{0}t}}{\sqrt{\nu_{0}t}}-\mathrm{e}^{-\nu_{0}t}\right)$,
where we have chosen $\nu_{\mathrm{0}}t_{\mathrm{tr}}=5\times10^{5}$.
The exponents of the power-law regimes are indicated with arrows.
Such regimes, especially at late times, can be indicative of dispersive
transport.}
\end{figure}

As was done in Fig. \ref{fig:current} for normal transport, Fig.
 \ref{fig:currentFractional} explores the effect that varying free
and trapped particle recombination rates has on time-of-flight current
transients by plotting the current given by Eq. \eqref{eq:tofCurrent}
for dispersive transport. For this, we have chosen to use the heavy-tailed
trapping time distribution derived in \citep{Philippa2014}: 
\begin{equation}
\phi\left(t\right)=\alpha\nu_{0}\left(\nu_{0}t\right)^{-\alpha-1}\gamma\left(\alpha+1,\nu_{0}t\right),\label{eq:multipleTrapping}
\end{equation}
where $\gamma\left(a,z\right)\equiv\int_{0}^{z}\mathrm{d}\zeta\,\zeta^{a-1}\mathrm{e}^{-\zeta}$
is the lower incomplete Gamma function and $\nu_{0}$ is a frequency
characterising the rate of escape from traps. In this case, the trap
severity has a physical interpretation as the ratio $\alpha\equiv T/T_{\mathrm{c}}$,
where $T$ is the temperature and $T_{\mathrm{c}}$ is a characteristic
temperature that describes the width of the density of states. In
Fig. \ref{fig:currentFractional} we use the same system of units
as Fig. \ref{fig:current} and all the same relevant parameters, except
for the trapping frequency which we increase to $\nu_{\mathrm{trap}}t_{\mathrm{tr}}=10^{4}$.
The new parameters that we must specify here are chosen as $\alpha=1/2$
and $\nu_{\mathrm{0}}t_{\mathrm{tr}}=5\times10^{5}$.

In Fig. \ref{fig:currentFractional}, the recombination-free current
transient is included in black as a reference. The most notable aspect
of this curve are the two power-law regimes indicative of dispersive
transport. The first power-law regime is analogous to the plateau
in Fig. \ref{fig:current}, as we have trapping and detrapping simultaneously
and contrarily affecting the current. However, unlike Fig. \ref{fig:current},
detrapping is such a rare event that we never reach a transient equilibrium
and the current decreases overall. The second power-law regime is
analogous to the rapid drop in current seen in Fig. \ref{fig:current}
after almost all free particles have been extracted. Here we actually
have a slower decrease in current as, unlike Fig. \ref{fig:current},
traps are so long-lived that detrapping events continue to contribute
to the current, even at very late times.

Fig. \ref{fig:currentFractional}a) considers an increasing free particle
recombination rate, $\nu_{\mathrm{loss}}^{\left(\mathrm{free}\right)}$.
Notably, as the free particle recombination rate increases, the first
power-law regime vanishes. In effect, the large recombination rate
of free particles causes an earlier emergence of the second power-law
regime that occurs when most free particles have been extracted. Thus,
it is also possible to conclude the existence of dispersive transport
from a time-of-flight current transient with a single power-law regime
at late times.

Fig. \ref{fig:currentFractional}b) considers an increasing trapped
particle recombination rate, $\nu_{\mathrm{loss}}^{\left(\mathrm{trap}\right)}$.
This subplot illustrates the necessity that there to be no trap-based
recombination for transport to be dispersive, as even a small amount
of trapped particle losses causes the second power-law regime to vanish.
We observe that the first power-law regime does not always vanish
completely and so it is important to note that the presence of a single
power-law regime at intermediate times does \textit{not} imply dispersive
transport.

\subsection{Ratio of particle detrapping to trapping, $R$, for fractional transport}

All of the results of the earlier sections depend in some way on the
steady-state ratio between particles leaving and entering traps, $R$,
defined explicitly in Eq. \eqref{eq:Rdefinition} or implicitly as
given by the integral in Eq. \eqref{eq:Rimplicit}. Unfortunately,
the latter integral definition is not expected to converge when fractional
transport is considered due to the asymptotic power law form \eqref{eq:heavyTail}
of the effective waiting time distribution. In this case, we have
the alternative definition:
\begin{equation}
R\equiv1+\frac{\Delta\nu_{\mathrm{loss}}}{\nu_{\mathrm{trap}}},
\end{equation}
valid irrespective of the chosen heavy-tailed trapping time distribution.
This definition provides an extension to the list of $R$ values in
Appendix A of \citep{Stokes2016} for fractional transport.

\subsection{Fractional Einstein relations}

The generalised Einstein relation \eqref{eq:generalisedEinsteinPara}
for diffusivity in the direction of the field can be simplified when
transport is fractional in nature. Here, as the mean trapping time
diverges, the average time $\tau$ defined by Eq. \eqref{eq:timeAverage}
also diverges, resulting in the fractional Einstein relation
\begin{equation}
\frac{D_{\parallel}}{K}=\frac{k_{\mathrm{B}}T_{\parallel}+mW^{2}}{e}\left[1+\left(1+\Delta\right)\frac{\mathrm{d}\ln K}{\mathrm{d}\ln E}\right],
\end{equation}
with
\begin{equation}
\Delta\equiv\frac{Q+\frac{3}{2}k_{\mathrm{B}}\left(T_{\mathrm{coll}}-T_{\mathrm{detrap}}\right)W\frac{\nu_{\mathrm{coll}}\left(\varepsilon\right)}{\nu_{\mathrm{eff}}\left(\varepsilon\right)}}{2k_{\mathrm{B}}T_{\parallel}W+2mW^{3}}.
\end{equation}
This fractional Einstein relation is valid for any trapping time distribution
with the asymptotic power law form of Eq. \eqref{eq:heavyTail}.

\section{\label{sec:Conclusion}Conclusion}

We have explored a generalised phase-space model that considers collision,
trapping, detrapping and recombination processes, all of which act
selectively according to particle energy. We form balance equations
\eqref{eq:continuityBalance}\textendash \eqref{eq:energyBalance}
describing the conservation and transport of particle number, momentum
and energy, and use these balance equations to form expressions for
the particle mobility, Eqs. \eqref{eq:zerothMobility} and \eqref{eq:firstMobility},
and for the average particle energy in the form of Wannier energy
relations \eqref{eq:zerothWannier} and \eqref{eq:firstWannier}.
These Wannier energy relations were then used to provide conditions
for particle heating or cooling due to collisions or trapping, Eq.
\eqref{eq:collisionCondition}, and recombination, Eqs. \eqref{eq:recombinationCondition1}
and \eqref{eq:recombinationCondition2}. Notably, recombination heating
and cooling was found to occur even when particles recombined indiscriminate
of energy, in contrast to the case where recombination occurs only
in the delocalised states. Transport via combined localised/delocalised
states was shown to produce negative differential conductivity under
certain conditions \eqref{eq:ndcCondition}, and the impact of scattering,
trapping/detrapping and recombination on the anisotropic nature of
diffusion was expressed through the development of the generalised
Einstein relations \eqref{eq:generalisedEinsteinPerp} and \eqref{eq:generalisedEinsteinPara}.
Lastly, fractional transport analogues of the aforementioned results
were explored by using a trapping time distribution with a heavy tail
of the form of Eq. \eqref{eq:heavyTail}.

For direct application of this model, it is necessary to have reasonable
inputs for the trapping frequency, $\nu_{\mathrm{trap}}$, and the
trapping time distribution, $\phi\left(t\right)$. Some progress has
been made already for organic materials where the trapping time distribution
can be calculated from the density of existing trapped states \citep{Philippa2014}.
Also for dense gases/liquids, where trapped states are formed by the
electron itself and the trapping time distribution is dependent on
the scattering, fluctuation profiles and subsequent fluid bubble evolution
\citep{Cocks2016a}. Other investigations of trapping also exist in
the literature \citep{Ceperley1980,Miller2008,Cao1993}, including
free energy changes and solvation time scales, but none of these directly
produces an energy-dependent trapping frequency or trapping time distribution.
Presently, the focus of our attention is on the \textit{ab initio}
calculation of energy-dependent trapping frequencies and waiting time
distributions in liquids and dense gases, as well as the simulation
of charge carrier transport in 2D organic devices, including those
with long-lived traps where transport is dispersive.

\appendix

\section{\label{sec:HigherMoments}Approximating higher order velocity moments}

In Sec. \ref{sec:BalanceFirst}, we use first-order momentum transfer
theory to obtain expressions for the drift velocity, Eq. \eqref{eq:firstMobility},
and mean energy, Eq. \eqref{eq:firstWannier}, of charged particles
defined by the generalised Boltzmann equation \eqref{eq:boltzmannEquation}.
These velocity moments are each expressed in terms of the higher order
velocity moments of energy flux $\boldsymbol{\xi}\equiv\left\langle \epsilon\mathbf{v}\right\rangle ^{\left(0\right)}$
and mean squared energy $\left\langle \epsilon^{2}\right\rangle ^{\left(0\right)}$.
Here, we use zeroth-order momentum transfer theory to approximate
these higher order moments by using the lower order ones.

In our previous work \citep{Stokes2016}, we consider constant process
rates in the Boltzmann equation \eqref{eq:boltzmannEquation}. This
is functionally equivalent to the case of zeroth-order momentum transfer
theory, as defined in Eq. \eqref{eq:constantEnergy}. In Eq. (74)
of \citep{Stokes2016} we write the solution of the Boltzmann equation
as a Chapman-Enskog expansion in Fourier-transformed velocity space.
By considering the first term of this expansion, we find an approximation
to the solution that is valid near the steady, spatially uniform state:
\begin{equation}
f\left(t,\mathbf{r},\mathbf{v}\right)\approx n\left(t,\mathbf{r}\right)\left[\omega_{\mathrm{coll}}\left(\varepsilon\right)\hat{w}\left(\alpha_{\mathrm{coll}},\mathbf{v}\right)+\omega_{\mathrm{trap}}\left(\varepsilon\right)\hat{w}\left(\alpha_{\mathrm{detrap}},\mathbf{v}\right)\right],\label{eq:approxSolution}
\end{equation}
where the convex combination weights $\omega\left(\varepsilon\right)$
are defined in terms of collision and trapping frequencies by Eqs.
\eqref{eq:collWeight} and \eqref{eq:trapWeight}. Here, the separate
processes of collision scattering and detrapping have resulted in
a solution containing non-Maxwellian velocity distributions of the
form
\begin{equation}
\hat{w}\left(\alpha,\mathbf{v}\right)\equiv w\left(\alpha,v\right)\frac{\sqrt{\pi}}{\sqrt{2}\alpha W}\mathrm{erfcx}\left(\frac{1-\alpha\mathbf{v}\cdot\alpha\mathbf{W}}{\sqrt{2}\alpha W}\right),
\end{equation}
where $w\left(\alpha,v\right)$ is the Maxwellian velocity distribution
defined by Eq. \eqref{eq:maxwellianDefinition}, $\mathbf{W}$ is
the drift velocity from zeroth-order momentum transfer theory, defined
in Eq. \eqref{eq:driftVelocity}, and the scaled complementary error
function is defined as $\mathrm{erfcx}\left(z\right)\equiv\frac{2}{\sqrt{\pi}}\int_{z}^{\infty}\mathrm{d}\zeta\,\mathrm{e}^{z^{2}-\zeta^{2}}$.

As expected, taking velocity moments of this solution \eqref{eq:approxSolution}
reproduces the zeroth-order momentum transfer theory expressions for
drift velocity $\mathbf{W}$, Eq. \eqref{eq:zerothMobility}, and
mean energy $\varepsilon$, Eq. \eqref{eq:zerothWannier}. In the
same vein, we can find approximations for higher order velocity moments
written in terms of these lower order moments, $\mathbf{W}$ and $\varepsilon$.
For energy flux we find
\begin{equation}
\boldsymbol{\xi}\approx\left(\frac{5}{3}\varepsilon+\frac{4}{3}mW^{2}\right)\mathbf{W},
\end{equation}
and for mean squared energy:
\begin{equation}
\left\langle \epsilon^{2}\right\rangle ^{\left(0\right)}\approx\frac{5}{3}\left[\omega_{\mathrm{coll}}\left(\varepsilon\right)\varepsilon_{\mathrm{coll}}^{2}+\omega_{\mathrm{trap}}\left(\varepsilon\right)\varepsilon_{\mathrm{detrap}}^{2}\right]+\frac{13}{3}\left(mW^{2}\right)^{2},
\end{equation}
which is written in terms of the separate mean energies of $\tilde{w}\left(\alpha_{\mathrm{coll}},\mathbf{v}\right)$
and $\tilde{w}\left(\alpha_{\mathrm{detrap}},\mathbf{v}\right)$,
given respectively: 
\begin{eqnarray}
\varepsilon_{\mathrm{coll}} & \equiv & \frac{3}{2}k_{\mathrm{B}}T_{\mathrm{coll}}+mW^{2},\\
\varepsilon_{\mathrm{detrap}} & \equiv & \frac{3}{2}k_{\mathrm{B}}T_{\mathrm{detrap}}+mW^{2}.
\end{eqnarray}

\begin{acknowledgments}
The authors gratefully acknowledge the useful discussions with Prof.
Robert Robson, and the financial support of the Australian Research
Council.

PS is supported by an Australian Government Research Training Program
Scholarship.
\end{acknowledgments}

\bibliography{references}

\end{document}